\documentclass[sigconf,nonacm]{acmart}

\usepackage{tabularx}
\usepackage{array}
\usepackage{xspace}

\newcommand{\tbaseline}{\textsc{Baseline}\xspace}
\newcommand{\tguidedwriteover}{\textsc{Guided-Write-Over}\xspace}
\newcommand{\tparsons}{\textsc{Solve-Code-Puzzle}\xspace}
\newcommand{\tverify}{\textsc{Verify-and-Review}\xspace}
\newcommand{\tpseudo}{\textsc{Interactive-Pseudo-Code}\xspace}
\newcommand{\texplain}{\textsc{Explain-before-Usage}\xspace}
\newcommand{\tlead}{\textsc{Lead-and-Reveal}\xspace}
\newcommand{\ttrace}{\textsc{Trace-and-Predict}\xspace}

\definecolor{highlightcolor}{HTML}{000000} 

\usepackage{listings}
\usepackage{xcolor}
\usepackage{acmart-taps}

\aptLtoXcmd{}{}

\AtBeginDocument{%
  \providecommand\BibTeX{{%
    \normalfont B\kern-0.5em{\scshape i\kern-0.25em b}\kern-0.8em\TeX}}}

\copyrightyear{2025}
\acmYear{2025}
\acmDOI{XXXXXXX.XXXXXXX}
\acmISBN{xxx-x-xxxx-xxxx-x/xx/xx}

\acmPrice{}
\setcopyright{cc}
\begin{document}
\title[Design Space of Cognitive Engagement Techniques with AI-Generated Code]{Exploring the Design Space of Cognitive Engagement Techniques with AI-Generated Code for Enhanced Learning}


\author{Majeed Kazemitabaar}
\orcid{0000-0001-6118-7938}
\affiliation{%
  \institution{University of Toronto}
  \city{Toronto}
  \state{Ontario}
  \country{Canada}
}
\email{majeed@dgp.toronto.edu}

\author{Oliver Huang}
\orcid{0000-0001-6118-7938}
\affiliation{%
  \institution{Department of Computer Science, University of Toronto}
  \city{Toronto}
  \state{Ontario}
  \country{Canada}
}
\email{oliver@dgp.toronto.edu}

\author{Sangho Suh}
\orcid{0000-0001-6118-7938}
\affiliation{%
  \institution{Department of Computer Science, University of Toronto}
  \city{Toronto}
  \state{Ontario}
  \country{Canada}
}
\email{sangho@dgp.toronto.edu}

\author{Austin Z. Henley}
\orcid{0000-0002-0494-5373}
\affiliation{%
  \institution{Carnegie Mellon University}
  \city{Pittsburgh}
  \state{Pennsylvania}
  \country{United States}
}
\email{azhenley@cmu.edu}

\author{Tovi Grossman}
\orcid{0000-0002-0494-5373}
\affiliation{%
  \institution{Department of Computer Science, University of Toronto}
  \city{Toronto}
  \state{Ontario}
  \country{Canada}
}
\email{tovi@dgp.toronto.edu}

\renewcommand{\shortauthors}{Kazemitabaar et al.}

\begin{abstract}
Novice programmers are increasingly relying on Large Language Models (LLMs) to generate code for learning programming concepts. However, this interaction can lead to superficial engagement, giving learners an illusion of learning and hindering skill development. To address this issue, we conducted a systematic design exploration to develop seven cognitive engagement techniques aimed at promoting deeper engagement with AI-generated code. In this paper, we describe our design process, the initial seven techniques and results from a between-subjects study (N=82). We then iteratively refined the top techniques and further evaluated them through a within-subjects study (N=42). We evaluate the friction each technique introduces, their effectiveness in helping learners apply concepts to isomorphic tasks without AI assistance, and their success in aligning learners' perceived and actual coding abilities. Ultimately, our results highlight the most effective technique: guiding learners through the step-by-step problem-solving process, where they engage in an interactive dialog with the AI, prompting what needs to be done at each stage before the corresponding code is revealed.
\end{abstract}



\begin{teaserfigure}
  \includegraphics[width=\textwidth]{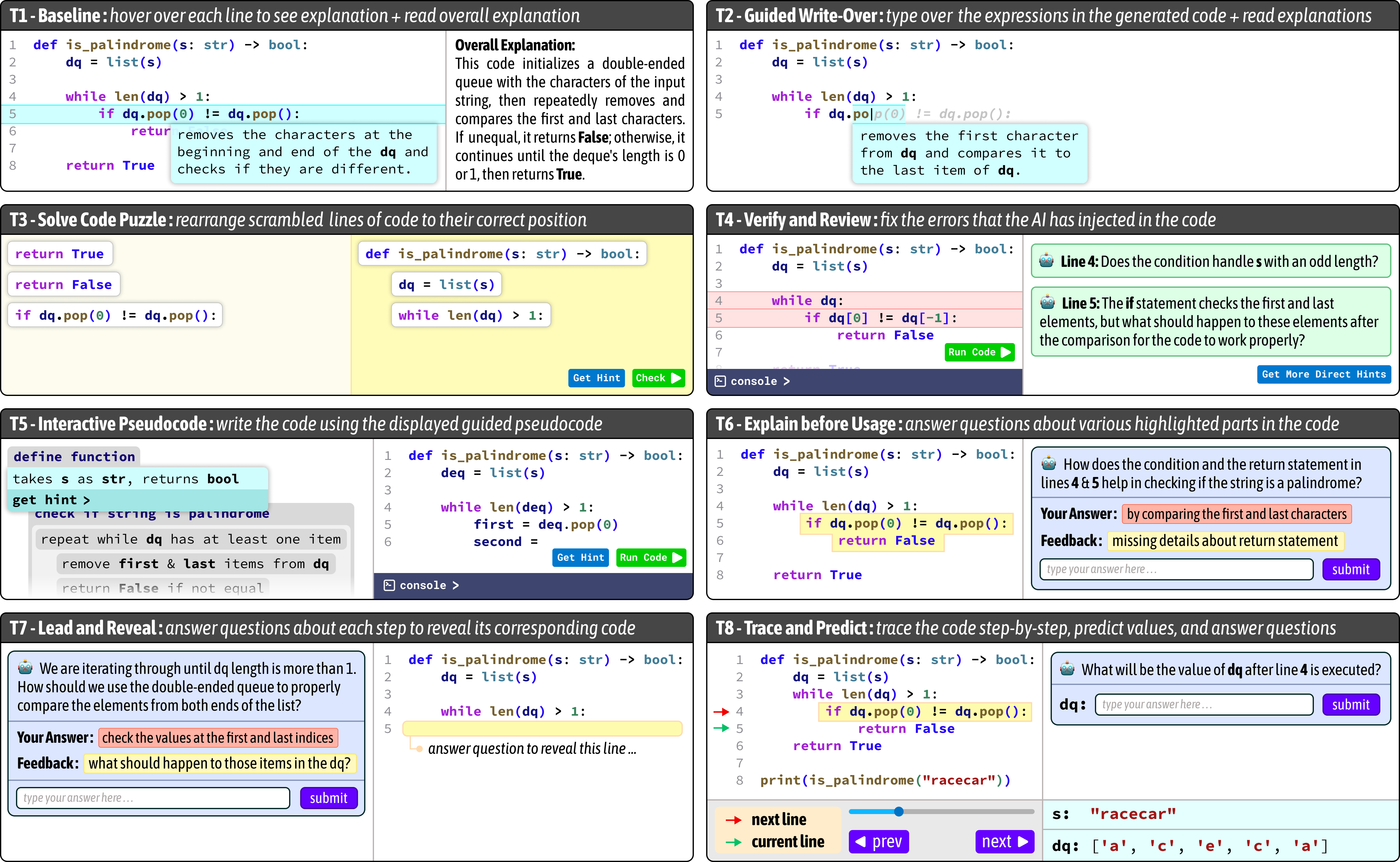}
  \caption{Illustration of the \tbaseline technique, displaying the AI-generated code and a comprehensive explanation, along with seven cognitive engagement techniques requiring varying levels of user engagement either before or after revealing the AI-generated code.}
  \label{fig:teaser}
\end{teaserfigure}

\maketitle

\section{Introduction}
Novice programmers often encounter challenges as they learn to code. Initially, learners struggle with understanding fundamental programming concepts and syntax \cite{qian2017intro_prog_difficulties, kaczmarczyk2010identifying, du2013some, bonar2013preprogramming, mayer1981psychology}, and as they advance, they face further difficulties mastering skills such as algorithmic thinking and software design principles \cite{hamouda2017basic, zingaro2018identifying, ragonis2005long, miedema2022identifying}.

Recently, Large Language Models (LLMs) are becoming increasingly integrated into novice programmers' workflow, influencing their help-seeking behaviors \cite{hou2024effects}, and changing the perspectives of both students and educators \cite{zastudil2023generative, lau2023ban}. Novice programmers now frequently use LLMs for various tasks, such as writing and explaining code, debugging, understanding concepts, and obtain example solutions \cite{zastudil2023generative, ghimire2024coding, hou2024effects}.

One prominent use of LLMs is for code generation \cite{ghimire2024coding} as demonstrated by large-scale self-report studies \cite{scholl2024novice} and analyses of student prompts \cite{scholl2024analyzing, kazemitabaar_koli23_using_ai}. Learners often cite the ability to frame questions easily and obtain faster, contextually relevant solutions as key advantages of using LLMs over traditional methods like web searches \cite{kazemitabaar_chi24_codeaid}. 

However, traditionally, learners proactively searched for and found relevant code examples to address gaps in their knowledge, meaningfully engaging themselves in a learning process \cite{brandt2009opportunistic, lahtinen2005difficulties, chatterjee2020novice_se_stackoverflow, skripchuk2023novice_web_help}. But with LLMs, learners can bypass the effortful yet beneficial process of adapting generic code examples to specific contexts \cite{ichinco2015exploring, wang2021learn_barriers_code_example}. While this shift can boost productivity, it can lower their engagement with the learning process and hinder the development of critical programming and Computational Thinking (CT) skills \cite{wing2006computational_thinking}. 

Currently, there are many concerns from both learners and educators about over-reliance on AI-generated code which could lead to skill degradation \cite{lau2023ban, zastudil2023generative, yilmaz2023augmented}. Learners might accept generated code without fully understanding it, giving them the illusion of learning \cite{prather2024benefits_harms_genai}. This can negatively impact their ability to write, modify, or debug code independently without AI \cite{kazemitabaar_koli23_using_ai, bastani2024genAi_harm_learning}. 

To address these challenges, prior research has sought to develop new learning activities to enhance prompting and code comprehension skills \cite{denny2024prompt_problems, denny2024explain_code_purpose}, or coding assistants that do not generate direct code solutions \cite{Liffiton2024codehelp} and instead, generate pseudocode \cite{kazemitabaar_chi24_codeaid}. However, given the efficiency and ubiquity of LLMs, it is likely that learners will continue to use unrestricted LLMs for code generation. Therefore, to prevent skill degradation, the focus should shift toward fostering deeper cognitive engagement with AI-generated code, rather than discouraging or limiting its use. Indeed, researchers are increasingly advocating for AI tools that promote metacognitive reflection \cite{tankelevitch2024metacognitive_demands} as a way to augment human cognition.

Engaging more deeply with AI-generated code involves critically analyzing it, evaluating the decisions that were made to solve the problem. This process involves rejecting information that contradicts their existing assumptions and adjusting their own knowledge to incorporate new insights. To promote such behavior, we drew inspiration from the concept of in-the-moment cognitive interventions \cite{park2019slow_algorithm, buccinca2021trust}, which have shown to enhance cognitive engagement and incidental learning \cite{gajos2022incidentallearning}. Our approach is to introduce a small amount of forced engagement with the AI-generated code---which we conceptualize as ``friction''---before learners can use the code. 

However, this friction must be carefully designed---it should not cause frustration or be seen as an unnecessary interaction. Instead, it should be thoughtfully designed to engage learners and provide meaningful assistance to their learning by, for example, helping them identify and bridge their knowledge gaps. Ideally, it would empower learners to not only understand the AI-generated code better but also to apply learned concepts to writing, extending, or modifying similar code.

Therefore, we systematically explored a broad design space of various forms of user engagements with code (Section \ref{sec:cogn_eng_techniques}). We developed seven distinct cognitive engagement techniques, illustrated in Figure \ref{fig:teaser}. Each technique introduces novel and unique forms of friction, requiring different types and levels of engagement. Another key factor is \textit{Engagement Timing}---whether to require user engagement before or after revealing the code solution. All techniques are designed based on the capabilities of LLMs and informed by prior research about supporting novice programmers in learning to code.

\begin{itemize}
    \item \textbf{T1 - \textit{\tbaseline}:} Users can directly use the generated code and accompanying explanation without any required engagement.
    \item \textbf{T2 - \textit{\tguidedwriteover}:} Users must type over each line of the generated code while being shown explanations about the purpose of each expression.
    \item \textbf{T3 - \textit{\tparsons}:} Users must reorder scrambled lines of AI-generated code into their correct horizontal and vertical spots.
    \item \textbf{T4 - \textit{\tverify}:} AI injects errors into the generated code and users must identify and fix them with guided support.
    \item \textbf{T5 - \textit{\tpseudo}:} AI generates pseudocode of the solution that users must manually implement with guided support.
    \item \textbf{T6 - \textit{\texplain}:} Users must answer AI-generated questions about various parts of the code.
    \item \textbf{T7 - \textit{\tlead}:} The AI guides users through solving the problem step-by-step, prompting them to explain the necessary actions and decisions for each step before revealing the corresponding line of code.
    \item \textbf{T8 - \textit{\ttrace}:} Users must trace the code execution line by line with a sample input and predict value of variables at key steps.
\end{itemize}

To evaluate the effectiveness of each technique, we conducted two studies with novice programmers recruited from intermediate-level computer science programming courses. From the first study (n=82) we identified the \tlead and \ttrace techniques as the most balanced techniques, fostering learning while eliciting just the right amount of cognitive engagement---enough to induce learning, but not so much as to overwhelm them. We then refined these two techniques based on collected feedback (Section \ref{sec:final_iteration}). Afterwards, we conducted a second, within-subjects study (n=42). The second study evaluated an additional metric: how effectively do techniques help users align their perceived and actual ability to write code of similar complexity without AI assistance? Our results showed that the \tlead technique better aligned participants' perceived and actual ability compared to the \tbaseline and \ttrace techniques, without increasing cognitive load (Section \ref{sec:second_study}).

Lastly, we synthesize our results and discuss the concept of ``\textit{Friction-Induced AI}'' to enhance short-term productivity gains while preventing long-term productivity loss due to over-reliance on AI.

This paper makes the following contributions:
\begin{itemize}
    \item \textbf{Design Space for Cognitive Engagement with AI-generated Code:} seven novel AI-assisted techniques that vary in engagement level, systematically positioned within the design space (Section \ref{sec:cogn_eng_techniques}).
    \item \textbf{Empirical Evaluation:} In a between-subjects study (n=82), we compared the seven techniques with the \tbaseline and identify \tlead and \ttrace as the most effective techniques for balancing learning gains and friction:  (Section \ref{sec:first_study}).
    \item \textbf{Refinement and Validation:} A within-subjects study (n=42) with the updated \tlead and \ttrace techniques shows \tlead significantly improves alignment between perceived and actual coding ability without increasing cognitive load (Section \ref{sec:second_study}).
    \item \textbf{Open-Source Release:} We provide the final \tlead technique as a new tool for education use and  future research: https://lead-and-reveal.vercel.app
\end{itemize}

\section{Related Work}\label{sec:related_work}
This section discusses the challenges faced by novice programmers, established interventions that address these difficulties, and the evolving role of Large Language Models (LLMs) in programming education, including their potential impact on metacognitive skills and over-reliance on AI.

\subsection{Challenges in Learning to Code}

Despite the growing interest and accessibility, novice programmers face persistent challenges when they start learning to code \cite{Altadmri2015SIGCSE, Becker2016ITiCSE, Brown2014ICER, Ebrahimi1994IJHCS, Hristova2003SIGCSE, Jackson2005FIE, Spohrer1986CACM}. They must simultaneously learn conceptual knowledge, syntactic knowledge, and strategic knowledge like planning, problem-solving, and debugging~\cite{qian2017intro_prog_difficulties}. End-user programmers, who write code for their own use, face six learning barriers, identified by Ko et al.: uncertainty about next steps, selecting appropriate programming constructs, combining constructs, using code construct correctly, understanding code failures, and inspecting program behavior~\cite{ko2004learning_barriers_eup}, with similar mistakes and misconceptions identified by other research \cite{Altadmri2015SIGCSE, Kolikant2005ICER, Kennedy2019ITiCSE, Lister2006ITICSE, McCauley2008CSE}. Furthermore, as novice programmers make progress, they face difficulties in acquiring advanced skills like algorithmic thinking and software design \cite{hamouda2017basic, zingaro2018identifying, ragonis2005long, miedema2022identifying}.

To address these challenges, researchers have introduced and experimented with various pedagogical strategies and software tools. A common approach in introductory programming courses is practicing with traditional coding tasks, where students write programs from scratch. While beneficial, these tasks often lead to frustration due to difficulties with syntax and conceptual understanding.

Parsons problems were introduced as an alternative, requiring learners to arrange shuffled lines of code into the correct order, emphasizing logical reasoning over syntax and error correction~\cite{parsons2006parsons}. Studies show that Parsons problems take less time to complete than fixing buggy code or writing code from scratch, without sacrificing learning outcomes~\cite{ericson2017parsons}. Further research has investigated variations such as \textit{faded} \cite{weinman2021faded_parsons} and \textit{adaptive} Parsons problems~\cite{ericson2018adaptive_parsons} with similar results. Worked examples are another alternative, which guide learners step-by-step through solving a problem, helping them understand how expert programmers approach coding tasks. They have been found particularly effective in reducing the time needed to reach proficiency~\cite{morrison2015workedexamples}, leading to the development of interactive versions \cite{gaweda2020typing_exercises}. Tracing exercises, where learners must follow the execution steps in a program, are frequently used to build understanding of programs execution and develop debugging skills \cite{lopez2008tracing_writing, venables2009tracing_writing_followup, kumar2013tracing_writing}.

Similarly, visualization tools allow learners to observe program execution and state changes, making abstract concepts like recursion and memory more tangible and accessible~\cite{sorva2013program_vis}. A notable example is Python Tutor, which displays visualizations of the program's data structures at each step of the code~\cite{Guo2013SIGCSE}. Other tools include Omnicode, which displays a scatter plot matrix of all run-time values for every variable in the program~\cite{Kang2017UIST}, Theseus, which annotates functions in the code editor with the number of times it was called during the current execution~\cite{Lieber2014CHI}, and an extension that displays a small graph of how each variable changes over time during execution~\cite{Hoffswell2018CHI}.

Another approach involves generating content to assist learners, such as hints~\cite{Galenson2014ICSE, Head2015VLHCC, Ichinco2018IDC, Peddycord2014EDM}, examples~\cite{Ichinco2016VLHCC, Ichinco2017CHI, Oney2012CHI}, tutorials~\cite{Harms2013IDC}, and recommendations~\cite{Fast2014CHI, Hartmann2010CHI, Mujumdar2011CHI, Suzuki2017VLHCC, Watson2012ICWBL}. Enhanced error messages provide more informative feedback on syntax and runtime errors, though results on their effectiveness have been mixed~\cite{Becker2016SIGCSE, Denny2014ITiCSE, Flowers2004FIE, Nienaltowski2008SIGCSE, Pettit2017SIGCSE, Prather2017ICER}.

However, as AI tools become increasingly accessible and learners rely on them for generating code solutions, they risk missing out on practicing and mastering essential higher-order cognitive skills like computational thinking \cite{wing2006computational_thinking}. To address this, our work builds on established interventions introduced through programming education research to re-engage learners with AI-generated code. Specifically, as we further elaborate in Section \ref{sec:cogn_eng_techniques}, we draw inspiration from these prior interventions and the associated findings to develop cognitive engagement techniques designed to promote cognitive engagement, deeper learning, and avoid over-reliance on AI.

\subsection{The Changing Landscape with Generative AI}

The rise of generative AI, especially Large Language Models (LLMs) is transforming programming education. These models can generate code from natural language, explain code, correct errors in code, and offer code completions (e.g., Github Copilot \cite{github_copilot}). They are now widely accessible through tools like ChatGPT and Claude AI and can serve as personalized coding assistants and tutors \cite{guo2023six_opportunities_chatgpt}. Recent research demonstrates that OpenAI GPT-3 outperformed students in multiple computer science exams \cite{finnie2024ai_cs2_tasks} and GPT-4 passing exercises of three Python programming courses without human involvement \cite{savelka2023gpt4_pass_csed}. Even multi-modal AI models can solve visual tasks like Parsons problems with high accuracy \cite{hou2024ai_solve_parsons}.

Although AI-powered tools can arguably offer exciting promises including improved engagement and personalized feedback ~\cite{Kasneci2023ChatGPT}, they also present challenges that limit learning opportunities. One major concern is over-reliance on AI-generated solutions~\cite{prather2023robots, prather2024benefits_harms_genai}. For novice programmers, tools like Copilot or ChatGPT provide quick answers to coding problems, which limits the opportunity for learners to critically engage with the problem and develop solution strategies~\cite{Becker2023CodingWasHard}. Researchers are already providing empirical evidence on the effect of access to AI on learning outcomes and over-reliance. Kazemitabaar et al. conducted a controlled experiment with 69 high-school students with no prior Python experience, comparing learning outcomes between two groups: one with access to AI and one without, over seven sessions, followed by two evaluation sessions without AI. Their results showed that the AI group experienced less frustration, and improved learning outcomes for students with stronger prior conceptual skills. But a follow-up analysis revealed various types of over-reliance on AI. Particularly, when students relied on AI to solve tasks autonomously, their subsequent coding skills without AI were consistently diminished \cite{kazemitabaar_chi23_effect_ai}. 

One approach to address over-reliance has been to place guardrails around AI so that it would avoid generating direct code solutions. Bastani et al. compared three groups of students, one with AI, one without, and one with an AI that safeguards learning (using prompt engineering) in a high school math class with nearly a thousand students. Their results demonstrate that while access to the AI boosts performance, students who later lose access to it perform worse than those who never had access to AI--except for those who used AI with guardrails \cite{bastani2024genAi_harm_learning}. Therefore, researchers and educators have called out for the development of pedagogical AI tools \cite{lau2023ban} which has led to the development of coding assistants like CodeAid \cite{kazemitabaar_chi24_codeaid} and CodeHelp \cite{Liffiton2024codehelp} that avoid displaying direct code solutions. Although these solutions might help, they are common in limiting and discouraging the use of AI-generated code. However, generative AI tools are here to stay and many acknowledge that we have crossed a point of no return with AI integration in programming education, and therefore, we should embrace this new paradigm \cite{lau2023ban}. In our work, we take an alternative approach and instead promote learners in deep, cognitive engagement with AI-generated code.

\subsection{Cognitive Engagement with AI}

Over-reliance on AI is a challenge that is being observed in many domains and workflows beyond learning to code. For example novice designers often take on AI suggestions without putting the effort to explore the alternatives~\cite{pace2024novice_architects_genai, xiaotong2024jamplate}. This over-reliance stems from cognitive biases and tendencies. One prominent example is automation bias, which describes the human tendency to favor suggestions from automated systems, often ignoring contradictory information from non-automated sources, even when the latter is correct. Extensive research, such as studies involving doctors using clinical decision support systems, has demonstrated the pervasiveness of this bias~\cite{goddard2012automation}. Another contributing factor is cognitive offloading, which suggests that people rely on external aids---in this case, AI---in order to reduce their cognitive load~\cite{risko2016cognitive}. Together, these concepts---automation bias and cognitive offloading---form the motivation of our work, highlight the challenges in integrating AI into human workflows and the need for interventions that strike the right balance between engagement and cognitive load.

While this challenge has existed since the development of automated systems, the explosive adoption of AI in everyday life and work has amplified it into a widespread problem, affecting millions of people---not just a few experts using decision support systems. Thus, a growing number of researchers are recognizing the need for interventions that help users engage more deeply with AI. For example, Tankelevitch et al. advocated that all generative AI-powered systems should provide metacognitive support strategies, such as explainability and customizability, and suggested mechanisms for eliciting user reflections on both their own decisions and those of AI~\cite{tankelevitch2024metacognitive_demands}. Similarly, Gajos et al. conducted three experiments with nutrition-related decisions and found that providing AI explanations without direct recommendations led to users learning more, as users had to derive their own conclusions from the explanations rather than relying on an 'answer' from the AI~\cite{buccinca_cscw21_overreliance_ai, gajos2022incidentallearning}. Buccinca et al. tested three cognitive forcing functions---interventions that require users to explain why they accept or reject AI recommendations---and found that such interventions can help reduce over-reliance on AI. However, they also observed that people rated these designs less favorably compared to those where they might over-rely on AI, understanding again that it is important to carefully design these interventions~\cite{buccinca2021trust}. 

In programming education, Kazemitabaar et al. studied the issue of learners passively accepting AI-generated code by developing CodeAid \cite{kazemitabaar_chi24_codeaid}, which generates pseudo-code, or highlights incorrect lines with suggested fixes, rather than directly displaying code solutions. Similarly, Hou et al. developed CodeTailor \cite{hou2024codetailor}, which responds to learners' help requests by presenting code solutions as Parsons problems for them to solve, instead of directly providing the fixed solution. Their evaluation study showed how this approach improved learning outcomes compared to a baseline where generated code was directly provided.

Our work builds on these insights in the context of AI-assisted programming and explore new interventions that can help learners cognitively engage with AI-generated code to ensure they retain higher-order CT skills like problem-solving.

\section{Cognitive Engagement Techniques with AI-Generated Code}\label{sec:cogn_eng_techniques}
To properly explore the design space of cognitive engagement techniques, we developed a set of initial prototypes, each with distinct types of user involvement based on the capabilities of LLMs and prior literature in supporting novices in learning to code. We then iteratively refined each technique through design probe sessions with five CS educators.

Additionally, to guide our design, we constructed a design space based on three key dimensions relevant to our context. First, Bloom's taxonomy which classifies learning objectives by cognitive complexity \cite{bloom1956taxonomy}: \textit{Remember}, \textit{Understand}, \textit{Apply}, \textit{Analyze}, \textit{Evaluate}, and \textit{Create}. Specifically, we use definitions adapted for programming education \cite{thompson2008bloom_cs}. Second, the ICAP framework \cite{chi2014icap_framework}, which links active learning outcomes to four categories of engagement level: \textit{Interactive}, \textit{Constructive}, \textit{Active}, and \textit{Passive}. And third, engagement timing: whether to require engagement with the AI-generated code before (\textit{Engage\&Reveal}) or after revealing the code solution (\textit{Reveal\&Engage}).

\begin{figure}
    \centering
    \includegraphics[width=1\linewidth]{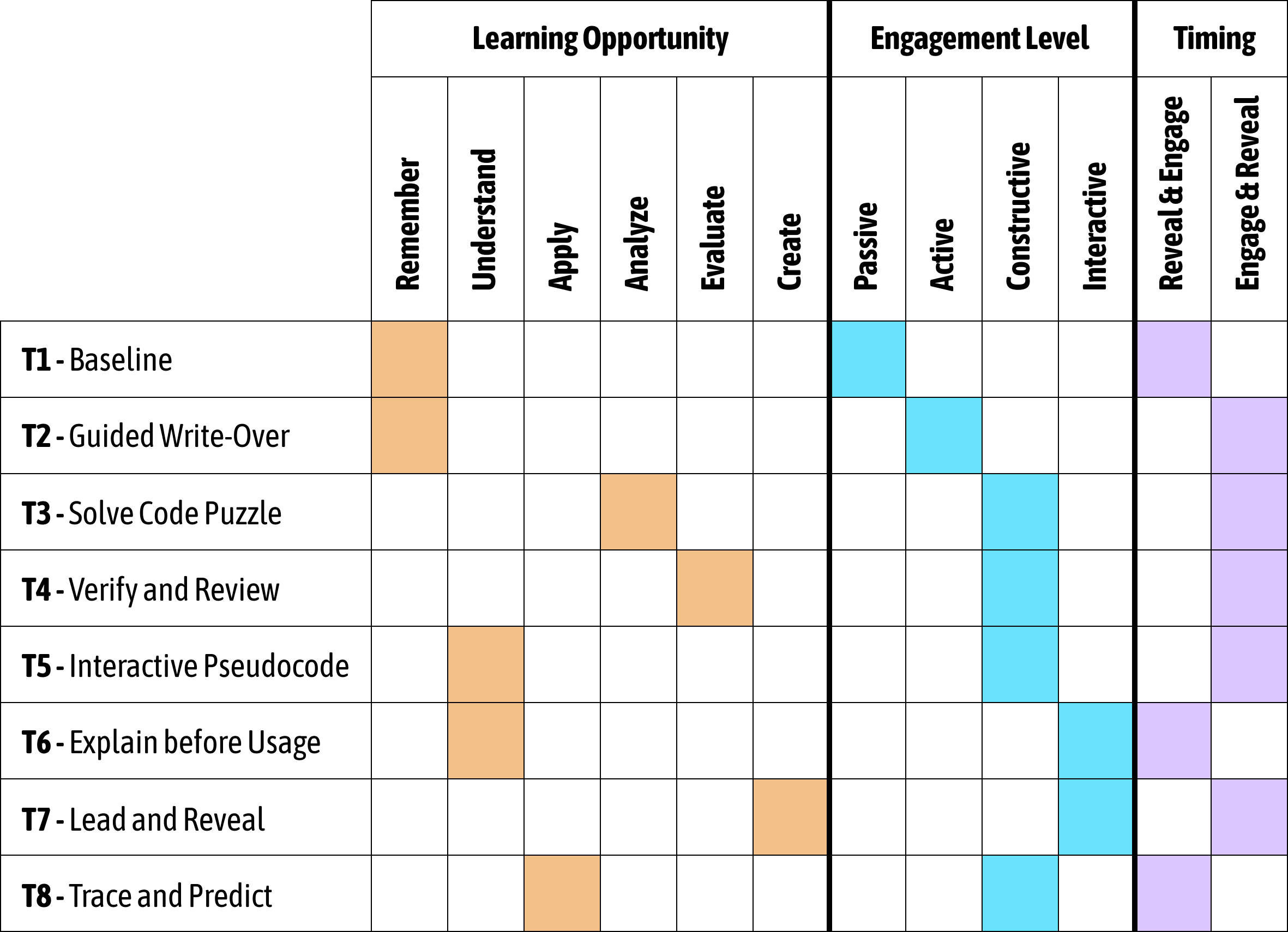}
    \caption{Summary of cognitive engagement techniques mapped within the design space. The first set of columns (in orange) represents learning opportunities aligned with Bloom's taxonomy. The second set (in blue) corresponds to cognitive engagement levels based on the ICAP framework. The final two columns (in purple) reflect whether the technique either reveals the code before requiring user engagement or requires user engagement before revealing the solution.}
    \label{fig:design_space_chart}
\end{figure}

We mapped each technique within the design space according to the above three dimensions, illustrated in Figure \ref{fig:design_space_chart}. We acknowledge that some degree of subjective interpretation is inherent in this process. Our design space is not meant to imply that these techniques can fit neatly into distinct areas. Rather, it was to guide us in making intentional design decisions. Below, we describe each technique, including the prior literature that inspired it and its design.



\subsection{T1. \tbaseline ( Minimal | Passive | Reveal \& Engage )}
In the \tbaseline technique, after the user provides the input prompt, the AI generates code and a detailed explanation using an initial prompt, and it is displayed to the user, similar to tools like ChatGPT. The level of engagement is completely optional and up to users. This technique induces little engagement, therefore classified as passive engagement.

\subsection{T2. \tguidedwriteover ( Remember | Active | Engage \& Reveal )}
For this technique, we drew inspiration from prior research in computing education, which found that requiring learners to type over worked examples improves learning~\cite{gaweda2020typing_exercises}. This technique increases learning opportunities by forcing deliberate practice \cite{ericsson1993deliberate_practice}, preventing passive copying of code without engaging one's attention to understand it.

In our technique, ghost text for each line of line of the generated code is displayed, similar to Github Copilot \cite{github_copilot}, but it requires users to type over it. A novel aspect of our technique is that, as users type, explanations for each expression (generated by a separate LLM prompt) are shown below the expression. Additionally, a high-level description of how the line contributes to the overall solution is displayed to the right of the line. Our technique also highlights incorrectly typed characters in yellow, prompting users to retype them. After two incorrect attempts, the system moves to the next character while marking the error in red. After the user successfully types over each line, they can proceed to use the code in their editor. 

This technique aligns with the "Remember" level of Bloom's taxonomy. It prompts users to actively retrieve previously learned concepts and recall relevant programming constructs, reinforcing memory retention and prevents shallow engagement.

\subsection{T3. \tparsons ( Analyze | Constructive | Engage \& Reveal )}
This technique adapts Parsons problem, which has been widely used in computing education for their demonstrated effectiveness. It requires learners to rearrange mixed code blocks to form a correct program \cite{parsons2006parsons, denny2008parsons}.
Empirical research has shown that various forms of these problems, including \textit{faded} \cite{weinman2021faded_parsons}, \textit{adaptive} \cite{ericson2018adaptive_parsons}, and \textit{personalized} \cite{hou2024codetailor}, can reduce cognitive load while maintaining learning performance \cite{ericson2017parsons}.


In this technique, each line of the AI-generated code is turned into a draggable block and is mixed inside a drag-and-drop pane, and the indentation is removed in the left pane. The user has to then drag each code block from the left, and put it in the correct spot on the right, as well as its correct level of indentation (horizontal spot). After using all code blocks, they can hit the "check" button that will only tell the user whether their solution is correct or not, and if all blocks are correctly placed they can proceed to use the code in their editor. There is also a "hint" button that will highlight all blocks that are placed incorrectly, and provide arrows to which direction each block should be moved to without explicitly telling them the exact position. 

This technique is positioned at the "Analyze" level as it requires analyzing what each block does and how it contributes to the solution, then they must rearrange them, "construct" sub-programs, and understand how the blocks work together to form a functioning program.

\subsection{T4. \tverify (Evaluate | Constructive | Engage \& Reveal )}
This technique draws inspiration from the use of debugging--a fundamental CT skill \cite{wing2006computational_thinking}--to help learners grasp programming concepts \cite{lee2014debug_learn_code, liu2017debugging_game, griffin2019bugs4learning}. Research beyond programming shows that incorporating incorrect solutions in worked examples can enhance learning transfer for advanced learners, but may overwhelm novices unless additional support, such as error highlighting, is provided \cite{groe2007fixing_worked_examples}. We applied these design guidelines to inform the design of the technique.

In this technique, the AI-generated code undergoes an additional LLM prompt that deliberately injects 3-4 errors, depending on the code's length and complexity. The user is then presented with this editable, incorrect code in a code editor, where they can run and debug it. The user is asked to identify and fix the injected errors. Furthermore, to assist users who might struggle with identifying the problems, clicking the "check" button highlights any remaining lines of code that needs to be fixed (using an LLM prompt). If they need help with how to fix the incorrect part, clicking on the "hint" button offers progressively more direct guidance, starting with hints about potential problems and eventually showing the correct code for each line. The user can use the editor to fix the problematic code and then check it with the "check" button. Once the code is error-free, the user can proceed to using the code in the editor.

This technique engages users at the "Evaluate" level by requiring them to actively check whether the code meets requirements or produces the correct output. Moreover, by fixing errors, they are required to critique the logic and quality of the code. They must determine which portions of the code are incorrect and fix those, while keeping the correct parts of the code.


\subsection{T5. \tpseudo (Understand | Constructive | Engage \& Reveal )}
We were inspired by the use of pseudocode in computing education to support algorithmic thinking before introducing complex syntax \cite{tew2011fcs1, linn1992case, robinson2016scratch_to_patch}, as well as by CodeAid, an LLM-powered assistant that encourages active learning by generating pseudocode rather than providing direct code solutions \cite{kazemitabaar_chi24_codeaid}.

This technique uses an additional LLM prompt to generate a subgoal-labeled, hierarchical pseudocode, displayed fully to the user without showing the code solution. The pseudocode is organized into subgoals, grouping several lines of code with a descriptive title. Users are provided with an empty code editor beside the pseudocode to write the corresponding code. Two buttons--``check'' and ``review''---use an LLM prompt to help guide the user by checking their code and offering feedback if errors are found. Initially, only subgoal titles are shown, with an "expand" button to reveal the pseudocode. Hovering over each pseudocode line provides a detailed explanation, and users can access two progressive levels of hints: syntactical guidance, followed by revealing the actual Python code. After successfully writing the code and verifying with the "check" button, users can proceed to using the code in their editor.

This technique reaches the ``Understand'' level as it helps them interpret the underlying logic and requires them to translate an abstract representation of the problem to concrete code, in a constructive manner.


\subsection{T6. \texplain ( Understand | Interactive | Reveal \& Engage )}
This technique is motivated by prior work on self-explanation, where learners were encouraged to explain the code. Researchers found that this can be a successful intervention, increasing engagement with worked examples \cite{vieira2017selfexplanation}, and that the ability to explain code has a significant relationship with the ability to write code \cite{lopez2008tracing_writing, venables2009tracing_writing_followup, murphy2012explain_writing_code}.

In this technique, after the AI generates the code, it is immediately displayed to the user. Immediately afterwards, an important part of the code (determined by a second LLM prompt) is then highlighted, followed by a short-answer question for the user to answer. The user's response is then evaluated via another LLM prompt with feedback and a score from 0 to 5, based on accuracy and completeness. The user has up to 3 attempts for each question, with the correct answer automatically shown after the third try. A total of 3-5 questions are asked from various parts of the generated code, depending on the code's complexity, after which the user can proceed to using the code in the editor.

This technique also reaches the "Understand" level as it engages users to interpret, explain, and infer meaning from existing code, without moving towards higher cognitive processes. However, it does this interactively, as users interact with AI to answer and receive feedback on their explanations.

\subsection{T7. \tlead ( Create | Interactive | Engage \& Reveal)}
This technique draws inspiration from CS Unplugged activities \cite{bell1998csunplugged}, which teach core concepts without any coding, and have been shown to improve computational thinking skills \cite{brackmann2017csunplugged}. We integrate this with scaffolded self-explanation using Socratic questioning, which has proven to enhance coding comprehension \cite{tamang2021socratic}.




After the code is generated, it is not displayed to the user. Instead, a second prompt is used to generate Socratic multiple-choice questions which are then displayed on one side of the interface. These questions scaffold the step-by-step process of solving the task and prompting the user in ``what needs to be done at each step of the algorithm/solution'' in a structured format. After the user picks the correct answer (where they have up to three retries), its corresponding line of code (which is the next line) will be displayed on the right with an explanation about how their answer and that line contributes to the overall solution. As each question is answered, its corresponding line of code will be progressively revealed on the right of the interface. This process is continued until all lines of the generated code is revealed and then the user can proceed to use the code in their editor.

This technique reaches the ``Create'' level as it engages users in creating a novel solution. They are prompted to interactively engage with the AI in hypothesizing and determining the next step in an algorithm. Since the code is not initially displayed, they are required to predict the logic, which stimulates creative thinking.


\subsection{T8. \ttrace ( Apply | Constructive | Reveal \& Engage )}
This technique is inspired by research showing that visualizing the notional machine is an effective method for learning programming concepts \cite{sorva2013program_vis, guo2013pythontutor}. Particularly, several studies have found a positive correlation between learners' performance on code tracing tasks and their code writing tasks  \cite{lopez2008tracing_writing, venables2009tracing_writing_followup, kumar2013tracing_writing}. Based on these findings, we aimed to strengthen learners' understanding by engaging them in code tracing tasks with integrated questions to sustain their engagement.

In this technique, the AI-generated code is displayed to the user with a sample input for execution. The user is then required to trace the code in a debugger-like system that includes buttons to step forward or backward through the code execution, and real-time variable values value of variables as they change. At certain complex parts of the code (identified by an LLM prompt), the technique does not allow the user to proceed. It requires the user to predict the value of a variable after that line of code is executed. The user can retry each variable prediction question three times and on the third try, the answer is displayed. This process is continued until several important parts of the code are covered with value prediction questions. After the user finishes tracing the entire code, and answering all the prediction questions along the way, they can then proceed to using the code in their editor.

Lastly, this technique reaches the ``Apply'' level as it requires users to examine the control flow, understand variable states, and identify how code executes line by line. Users have to actively solve questions about unfamiliar code, applying the rules of execution in a guided environment.

\section{Study 1: Comparing All Techniques}\label{sec:first_study}
We conducted a between-subjects study with 82 participants to evaluate each technique against the \tbaseline technique without forced cognitive engagement. Our goal was to evaluate:
\begin{itemize}
    \item \textbf{RQ1 [Performance]:} How effective are the cognitive engagement techniques in supporting participants' ability to transfer learned programming concepts to isomorphic coding tasks without AI assistance?
    \item \textbf{RQ2 [Friction]:} How do the techniques impact perceived friction, measured through time on task and cognitive load?
    \item \textbf{RQ3 [Perceptions]:} What are participants' perceptions about technique and the type of user involvement that they require?
\end{itemize}

\subsection{Experiment Tool Design}~\label{sec:experiment_tool}
To evaluate the effectiveness of our techniques, we built a web-based application that allows users to log in and complete AI-assisted or manual coding tasks. This also enabled specific tasks to be assigned to each user by the experimenter. The tool was designed to provide a self-paced and consistent experiment for all participants. In AI-assisted tasks, users interact with a chatbot to generate code. To avoid generating incorrect or unrelated code, the chatbot compared the user's prompt with the current task and provided feedback if details were missing. Additionally, to ensure consistent outputs across participants, pre-generated responses for each task and technique were used. In the \tbaseline technique, code is immediately shown with an accompanying explanation. For other techniques, a modal that covers the entire screen displays the technique after a 5-second delay to simulate generation time. Once users finish interacting with the technique, the modal disappears and users can see the code and its explanation.

The tool is developed using TypeScript with a client-server architecture. The server, implemented in Node.js, stores user data and logs using a MongoDB database, and interacts with OpenAI APIs (GPT-4). It parses LLM outputs and communicates with a Python shell and language server for running Python code and providing real-time autocomplete and error-detection for the client-side code editor. The Monaco code editor is used for code editing, execution, and submission. The user interfaces can be seen in \ref{fig:teaser}. Full details, including prompts and source code, are available on GitHub: https://github.com/MajeedKazemi/code-engagement-techniques.

\subsection{Methodology}
\begin{figure*}
    \centering
    \includegraphics[width=1\linewidth]{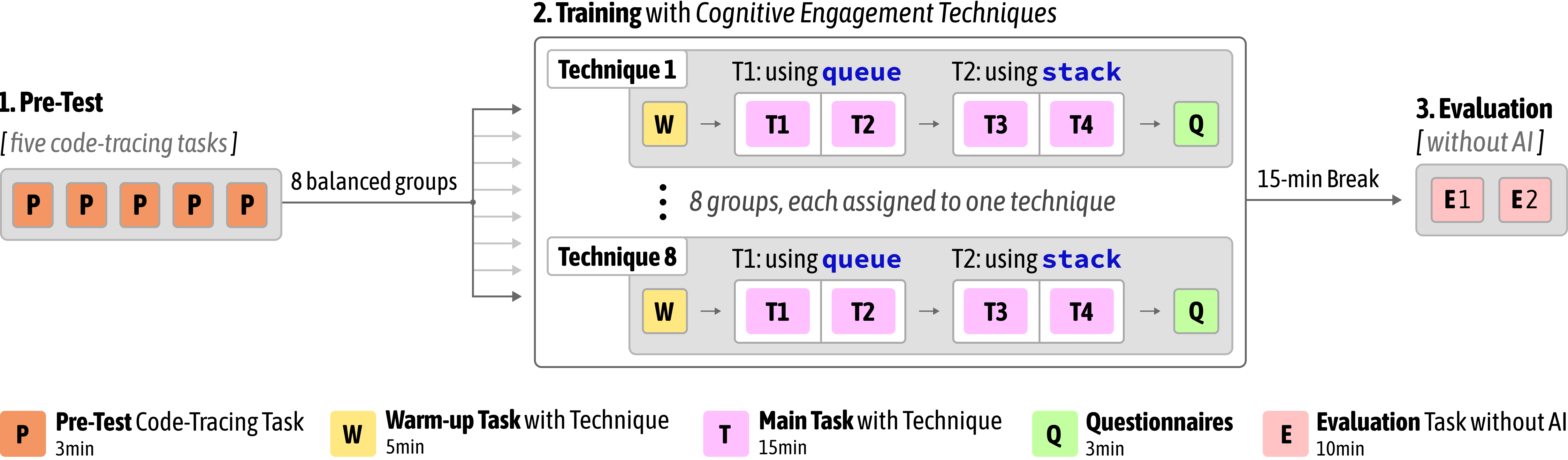}
    \caption{Overview of Study 1 procedure: Participants began with a pre-test, were split into 8 balanced groups, trained with AI using their assigned cognitive engagement technique, completed post-training questionnaires, took a short break, and then performed manual coding tasks for evaluation.}
    \label{fig:study1_procedure}
\end{figure*}

The study design posed two primary challenges: (1) ensuring participants had similar backgrounds to minimize variance in prior knowledge, and (2) designing programming tasks that would uniformly challenge participants, requiring assistance and enabling both learning and the transfer of learned concepts to isomorphic tasks. The following sections outline the methodology used to address these challenges. An overview of the study procedure is displayed in Figure \ref{fig:study1_procedure}.

\subsubsection{Participants}
We recruited 82 participants (44 male, 37 female, 1 non-binary), ages 18--23 ($M=19$), from an undergraduate programming course about data structures at a large public university. To ensure a consistent level of knowledge, all participants were recruited from the same class. Participants reported their frequency of using AI tools like ChatGPT for programming as daily (n=14), weekly (n=33), monthly (n=8), rarely (n=22), and never (n=5). Additionally, they reported using ChatGPT primarily for fixing code (n=59), explaining concepts (n=54), providing code snippet explanations (n=47), and generating code from descriptions (n=32). The study was approved at our institutes ethics review board, and informed consent was obtained from all participants prior to the study. Each participant received \$25 as compensation.

\subsubsection{Study Procedure} As shown in Fig.~\ref{fig:study1_procedure}, our study procedure consisted of three phases: pre-test, training, and evaluation. We describe each phase in detail below.

\begin{itemize}
    \item \textbf{Phase 1. [Pre-Test] 5 Code-Tracing Questions.} All participants initially attended an online 30-minute session over MS Teams to perform a pre-test, including five multiple-choice code tracing questions about the stack and queue data structures. The scores were then used to create eight balanced groups, each being randomly assigned to one of the seven experimental or the \tbaseline techniques. The mean pre-test scores ranged from 43.5\% to 49.0\%, with a standard deviation of 2.9\% across the group means. Each group had 10 participants, except for the \textit{\tverify} and \textit{\tbaseline} groups, which included 11 participants.
    \item \textbf{Phase 2. [Training] 4 Tasks with Assigned Cognitive Engagement Technique.} Participants then attended an in-person 2-hour session (including a 10-minute break) for the main study in which they used the web-based experiment tool. Participants logged in with their given credentials, where they were automatically assigned to the technique determined in advance. They then were instructed about how to use their assigned technique, followed by a warm-up task using the technique, and four training tasks with their technique on slightly complex tasks: two using stacks, and two using queues.
    
    The tasks were designed to be slightly outside of participants' zone of proximal development \cite{vygotsky1978zone}. We consulted with the instructor of the course from which participants were recruited to ensure that the tasks were useful for their learning, were sufficiently challenging to solve independently, and could potentially be learned through sufficient, deep engagement with the code solution. 
    
    Each task in our experiment tool first displayed the task description, several test-cases, and a multiple-choice question that asked participants how well they understand the task as a way to prompt them and to make sure they understood the task before starting to work on it. While working on the AI-assisted tasks, participants were given the option to either copy the task description as provided or rephrase it in their own words. Additionally, since engaging with the AI-generated code beyond the requirements of each technique was optional, participants were advised to move on to the next task once they felt confident in their understanding, without exceeding 15 minutes per task.

    \item \textbf{Phase 3. [Evaluation] Survey + 2 Coding Tasks without AI Assisance.} After completing the four training tasks, participants were asked to first fill out a survey, followed by working on two manual coding tasks without AI assistance.
    
    The survey included questions about perceived learning after using each technique, their perceived task load evaluated using the NASA Task Load Index (TLX) \cite{kosch2023nasa_tlx}, and several Likert-questions about their willingness to use the techniques, and open-ended questions about what they liked and disliked about the process of using the technique that they were assigned to.
    
    Participants then proceeded to work on two evaluation tasks, one being isomorphic to the second task about stacks and the other being isomorphic to the fourth training task on queues, with minimal changes in task requirements. No starter code was provided for the evaluation tasks. A timer was displayed on top of these tasks and participants were instructed to skip the task if they were not able to finish the task within 20 minutes.
\end{itemize}

\subsubsection{Data Analysis}
Learners' responses to the manual coding tasks were graded using two detailed rubrics, each tailored to one of the two tasks. These rubrics focused on key concepts and the use of data structures and how they were used in the training tasks. After the first author tested and refined the rubric on 25\% of the responses, it was applied to all 82 submitted codes for each task by the first author. For each of the training tasks we also collected the time they spent on each task, which could be an indicator of how much friction the technique caused.

For statistical analysis, we used a generalized linear model (GLM) to examine the effect of the intervention techniques on learners' evaluation scores while controlling for pre-test scores. One-way ANOVA is used for normally distributed data with equal variances, while the Kruskal-Wallis H test is the non-parametric alternative for non-normal data. For pairwise comparisons, an independent t-test is applied for normally distributed data, and the Mann-Whitney U test for non-normal data. In this study, we focused on pairwise comparisons between each of the seven experimental techniques and the \tbaseline technique, yielding a total of seven comparisons. To control for the risk of Type I error due to multiple-comparisons, we applied a Bonferroni correction, setting the threshold for statistical significance at $\alpha = \frac{0.05}{7} = 0.007$.

\subsection{RQ1 [Performance] How effective are the cognitive engagement techniques?}\label{sec:study1_rq1_performance}
Regarding the effect of the techniques on learners' ability to write code manually without AI assistance, the generalized linear model explained 17.7\% of the variance in post-test manual coding scores (\textit{Pseudo} $R^2 = 0.177$). Pre-test scores were a significant predictor of post-test performance ($b=0.34$, $SE=0.12$, $z=2.78$, $p=.005$), indicating that higher pre-test scores were associated with higher post-test scores. Compared to the \tbaseline technique ($M=31.5$, $SD=28.2$) which involved no forced engagement and demonstrated nearly the lowest mean performance, the \tlead technique ($M=58.3$, $SD=39.8$) demonstrated the greatest improvement ($b=26.8$, $p=.058$), though this did not reach statistical significance. The second most effective technique was \tparsons ($M=54.1$, $SD=38.2$; $b=22.8$, $p=.108$), followed closely by \ttrace ($M=54.1$, $SD=37.9$; $b=21.8$, $p=.123$).

A post-hoc power analysis ($\eta^2 = 0.112$) showed the study was underpowered with 82 participants. To reach 90\% power, a sample size of 153 would be required, suggesting the non-significant findings may be due to insufficient power. Future studies should aim for a larger sample.

\begin{figure*}
    \centering
    \includegraphics[width=1\linewidth]{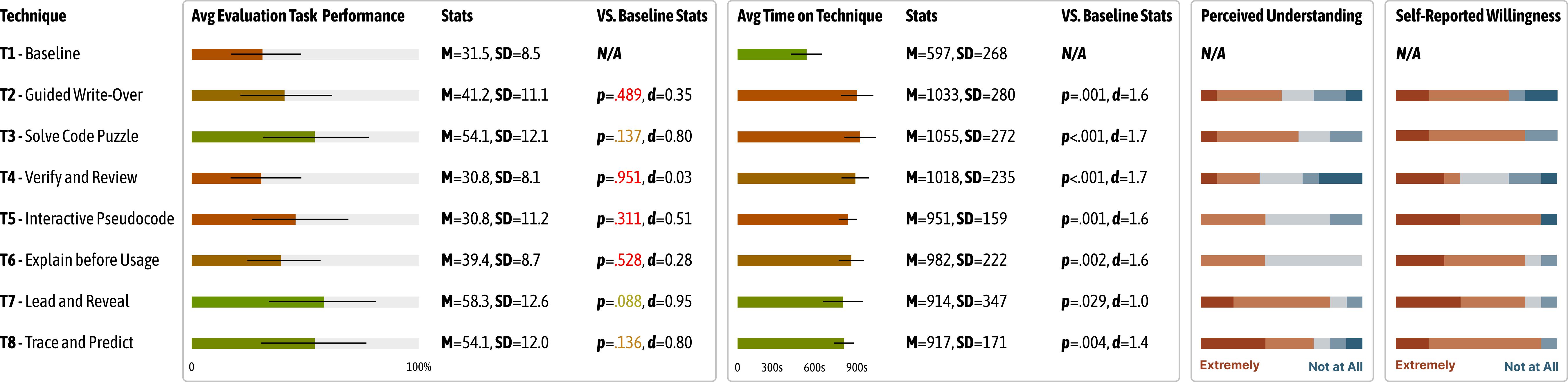}
    \caption{An overview of results from Study 1 (left to right): the performance on the two manual coding tasks. Average time spent on each technique during the training. Perceived understanding and reported willingness to use each technique (on 5-point Likert scales).}
    \label{fig:study1_results}
\end{figure*}

\subsection{RQ2 [Friction] How do the techniques impact perceived friction?}
To compare participants' perceived friction across techniques, we analyzed both task completion time and NASA Task Load Index (TLX) scores. A Kruskal-Wallis H test was used to examine differences in the six TLX dimensions across the eight techniques. Significant differences were found in frustration ($\chi^2(2, N=82)=19.5$, $p=.006$), physical demand ($\chi^2(2, N=82)=16.0$, $p=.025$), and effort ($\chi^2(2, N=82)=14.1$, $p=.05$). A Mann-Whitney U test showed that \tverify ($M=72.7$, $SD=19.6$; $p=.035$) had significantly higher frustration compared to the \tbaseline ($M=46.7$, $SD=27.9$). Higher physical demand was reported for \tpseudo ($M=60.0$, $SD=19.9$; $p=.005$) and \tguidedwriteover ($M=65.7$, $SD=28.7$; $p=.015$) compared to the \tbaseline ($M=36.4$, $SD=14.8$).

Task completion time was analyzed using a one-way ANOVA, which revealed significant differences ($F(7, 74)=3.64$, $p=.002$). The \tbaseline had the shortest time ($M=596s$, $SD=268s$), while \tparsons ($M=1055s$, $SD=272s$; $t(19)=3.89$, $p<.001$, $d=1.7$) and \tguidedwriteover ($M=1033s$, $SD=280s$; $t(19)=3.64$, $p=.001$, $d=1.6$) took the longest. \tlead ($M=914s$, $SD=347s$) was closest to the \tbaseline, with the smallest effect size difference ($t(19)=2.36$, $p=.029$, $d=1.0$).

Lastly, the number of times that participants tested and executed the AI-generated code after they finished the technique was statistically significant, indicated by a Kruskal-Wallis H test ($\chi^2(2, N=82)=23.16$, $p=.001$). The \tbaseline technique had the most number of runs ($M=3.9$, $SD=4.7$), while the \tpseudo ($M=0.3$, $SD=0.6$), followed by \ttrace ($M=0.6$, $SD=0.7$), and \tverify ($M=1.0$, $SD=1.6$) had the least. This could be explained by how the user has already executed the code and checked the output in the techniques before using the code in the editor. 

\subsection{RQ3 [Perceptions] What are participants' perceptions of the techniques?}

Below, we summarize participants' ($P1_{T1}$–$P10_{T8}$) perceptions of each technique based on their responses to open-ended questions about their likes, dislikes, and the impact on their understanding and engagement with AI-generated code.

\subsubsection{T1: \tbaseline}
Several participants acknowledged that the AI-generated code and explanations helped them grasp core concepts, with one noting that it facilitated understanding of the ``main concepts'' ($P7_{T1}$) and another stating it helped ``figure out the logic [they] needed'' ($P3_{T1}$). However, other participants indicated that this approach provided ``only a basic idea of how the code works'' ($P11_{T1}$). The majority emphasized that having the AI provide solutions directly did not foster meaningful learning. As $P10_{T1}$ remarked, ``directly giving [them] the answers was not helpful for learning,'' a sentiment echoed by others ($P1_{T1}$, $P2_{T1}$, $P5_{T1}$, $P6_{T1}$). For instance, $P5_{T1}$ reflected, ``without actually interacting much with the code the AI had written, I wasn't able to apply it.'' This feedback underscores the limited efficacy of passive AI-generated content for promoting deeper engagement and independent code-writing skills.


\begin{figure*}
    \centering
    \includegraphics[width=1\linewidth]{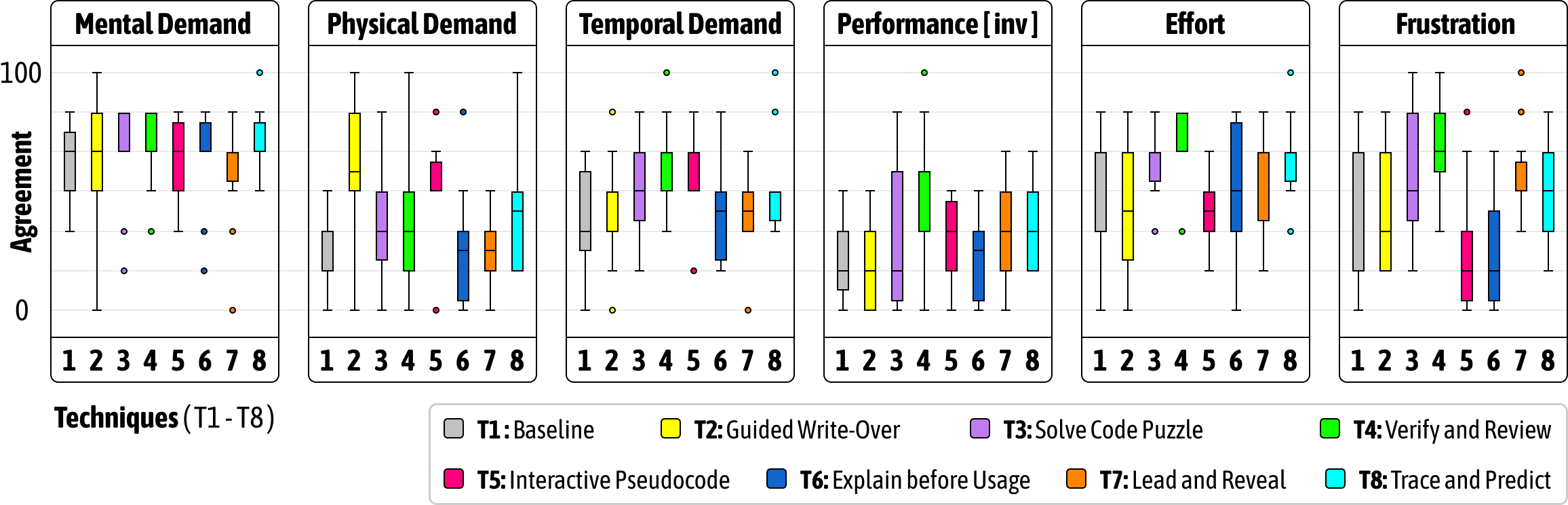}
    \caption{Cognitive load results for 8 techniques. Each box-plot displays the six dimensions of the raw NASA TLX (mental, physical, temporal demand, inverted performance, effort, and frustration) reported by participants after using each the techniques for four coding tasks. Each technique is represented by a different color, with higher values indicating higher task load.}
    \label{fig:study1_task_load}
\end{figure*}

\subsubsection{T2: \tguidedwriteover}
Participants generally appreciated the \tguidedwriteover technique for its ability to simulate the coding process, with several noting that it felt ``similar to writing the code [themselves]'' ($P8_{T2}$) and encouraged them to ``pay attention to the small details'' ($P6_{T2}$). They found the inline explanations beneficial for learning, particularly in understanding the purpose of individual expressions ($P4_{T2}$, $P6_{T2}$, $P9_{T2}$). However, some participants found the approach cumbersome, with one stating that it was ``tedious and slowed down [their] working pace'' ($P7_{T2}$) and others describing it ``simple and repetitive'' ($P10_{T2}$), and even ``extremely annoying'' ($P6_{T2}$). A recurring critique was that the explanations focused too heavily on \textit{what} the code did, neglecting the \textit{why} behind the solution in the broader context of problem-solving ($P3_{T2}$, $P8_{T2}$). Furthermore, there were concerns that the technique limited independent problem-solving, with one participant noting that it ``forces the user to follow [the AI's] algorithm, which may result in less practice in critical thinking and problem-solving'' ($P6_{T2}$), while another mentioned it ``did not address discrepancies'' between their own expected approach and the AI-generated code ($P7_{T2}$).

\subsubsection{T3: \tparsons}
Participants expressed that the process of actively assembling code fragments fostered deeper critical engagement with code structure and logic, with several noting it as ``a great learning tool'' ($P2_{T3}$) that ``forces you to have some understanding of how the solution works'' ($P4_{T3}$). Many appreciated the cognitive challenge, observing that it ``asks you to actually solve the problem'' ($P1_{T3}$) and ``think about how the function should be implemented'' ($P7_{T3}$). Some valued the structured breakdown of the task, which ``reduced stress'' and made the process feel ``more interesting, like playing with puzzles'' ($P5_{T3}$). However, the task also elicited mixed reactions, with a few participants describing it as ``mentally exhausting'' ($P6_{T3}$) and ``overwhelming'' ($P9_{T3}$), particularly when struggling to comprehend the purpose of individual code lines. This led to feelings of being ``mentally tired'' ($P6_{T3}$) and finding the task ``difficult to understand and learn'' ($P5_{T3}$). The rigid enforcement of code ordering further exacerbated frustration for some, as it ``required some of the code to be in an exact order even though it did not matter in some cases'' ($P7_{T3}$).

\subsubsection{T4: \tverify}
The technique elicited mixed reactions from participants, with several highlighting both its strengths and limitations. Some participants appreciated the active engagement required, with P1 noting that it ``forced me to manually process the code,'' and $P6_{T4}$ emphasizing that the progressive hint system ``contributed to my learning as it didn't give me the answer straight away.'' However, this benefit was tempered by concerns over the hint system fostering dependency, with $P9_{T4}$ remarking that the hints became ``straight-up corrections'' and $P10_{T4}$ explaining that this led to a lack of cognitive effort because they ``don't need to think.'' $P7_{T4}$ saw value in the technique's ability to ``enhance the ability to understand code written by others'' as it encouraged [them] ``to think about the kinds of mistakes that might occur in the code,'' yet this was counterbalanced by a more prevalent issue: the injected errors often caused confusion rather than promoting deeper learning, as expressed by $P2_{T4}$, $P3_{T4}$, $P4_{T4}$, $P5_{T4}$, $P7_{T4}$, and $P11_{T4}$. Additionally, the lack of debugging features in the code editor was a significant frustration for some participants, with $P1_{T4}$ noting that such features are ``integral to [their] understanding of code.'' Overall, while the technique showed promise in fostering active code comprehension, it also introduced usability challenges that hindered its effectiveness.

\subsubsection{T5: \tpseudo}
Participants expressed mixed reactions to the hierarchical pseudocode technique. Several appreciated how it encouraged problem-solving and independent thinking, with $P1_{T5}$ noting it pushed them ``to think the solution by [them]self,'' and $P2_{T5}$ appreciating the ``chance to try [them]self and implement the code based on the skeleton.'' The multi-level hint system was also praised, as $P6_{T5}$ valued ``how it doesn't immediately reveal the answer.'' However, $P8_{T5}$ felt they were ``just following instructions'' rather than truly creating code. While $P7_{T5}$ found the pseudocode helpful for ``break[ing] down a complicated problem into a digestible manner,'' others ($P2_{T5}$, $P9_{T5}$, $P10_{T5}$) found it unclear, especially for complex tasks. This highlights both the strengths and limitations of the approach in facilitating learning with AI-generated code.

\subsubsection{T6: \texplain}
Participants generally appreciated the technique's ability to enhance engagement and deepen their understanding of AI-generated code. Many noted that it ``helped [them] understand each crucial step in the code'' ($P1_{T6}$), ``internalize the purpose of each line'' ($P7_{T6}$), and ``prompted [them] to think deeper'' ($P5_{T6}$, $P6_{T6}$), while also being intellectually challenging ($P5_{T6}$). However, some expressed concerns that the technique ``only managed to explore surface level knowledge'' ($P4_{T6}$) and ``did not help [them] understand the code as a whole, just little bits of it'' ($P4_{T6}$), particularly when it ``would ask about a line of code whose purpose is rather straightforward'' ($P9_{T6}$). Participants appreciated the feedback provided, as it ``could point out my mistakes'' ($P1_{T6}$) and helped them identify ``why [their] logic was flawed'' ($P7_{T6}$), though some desired more personalized, detailed feedback rather than ``just showing them the correct answer'' ($P5_{T6}$).

\subsubsection{T7: \tlead}
Learners appreciated that the technique encouraged them to ``think of the purpose of each line of code'' ($P3_{T7}$) and helped them ``get [their] own idea of what needs to be done to solve it'' ($P5_{T7}$), with its sequential approach enabling them to ``break down the task into pieces'' ($P6_{T7}$). This scaffolded process guided learners to ``follow the correct thinking process of the code step by step'' ($P7_{T7}$). However, some participants found certain questions lacking ``comprehensive explanations and context'' ($P2_{T7}$), leading to confusion as they had to ``decipher why that approach was used'' ($P5_{T7}$), which at times ``altered [their] expectations for what [they] needed to do to solve the problem'' ($P1_{T7}$) when they had a different solution in mind. This resulted in participants like $P10_{T7}$ feeling they were ``constantly guessing what the AI was doing.'' Several participants ($P2_{T7}$, $P4_{T7}$, $P7_{T7}$) suggested that providing hints or explanations after incorrect answers would be beneficial in alleviating these challenges.

\subsubsection{T8: \ttrace}
Participants generally found the technique beneficial for fostering deeper engagement with AI-generated code. Many highlighted that tracing variable values ``helped [them] slow down and think more about the code'' ($P4_{T8}$), promoting a clearer understanding of ``each line of the code'' ($P2_{T8}$) and ``how the code works'' ($P5_{T8}$). The prediction questions encouraged users to engage critically, enhancing comprehension of ``concepts and strategies used to solve problems'' ($P3_{T8}$) and ``forcing [them] to think about what was happening'' ($P8_{T8}$), contributing to their learning ($P10_{T8}$). However, the technique's step-by-step process was seen as overly time-consuming and ``strenuous'' ($P10_{T8}$), with users expressing frustration over ``a lot of steps and questions'' ($P3_{T8}$) and the lack of a ``skip button'' ($P3_{T8}$), even for straightforward sections. Participants also disliked the absence of explanations for incorrect predictions or code design choices, stating that ``the system should have shown some explanation'' ($P10_{T8}$) and that there were ``no explanations as to why the code was designed in that way'' ($P1_{T8}$).

\subsection{Summary of Results}
Overall, the \tlead, \ttrace, and \tparsons techniques emerged as the most effective, showing improvements in learners' manual coding performance, although these differences did not reach statistical significance. Both \tlead and \ttrace had the highest performance overall, without imposing a meaningful task load and having the closest completion time to the \tbaseline. Conversely, techniques such as \tverify and \tguidedwriteover induced higher levels of frustration and physical demand, correlating with lower performance and perceived learning. While \tpseudo and \texplain showed some cognitive benefits, they were also seen as cognitively taxing and limited in promoting deeper problem-solving.

These results underscore a critical trade-off within the \textit{Engagement Timing} spectrum, where the final code solution is either revealed before user engagement or withheld until after engagement. When solutions are presented upfront, as in \ttrace or \texplain, users may feel overwhelmed, but this approach offers a comprehensive view of how each line of code contributes to the overall solution, potentially enhancing comprehension. Interestingly, techniques from both ends of the engagement spectrum---\tlead and \ttrace---showed similar improvements over the \tbaseline without causing excessive friction. This indicates that these contrasting approaches are equally valuable and warrant further investigation.



\section{Final Design Iteration}\label{sec:final_iteration}
To further explore the design space, we iterated on the \tlead and \ttrace techniques based on participant feedback from Study 1. We focused on these techniques as they effectively balanced performance and imposed friction. Additionally, We updated all techniques, including the \tbaseline, to provide line-by-line explanations on hover, inspired by \cite{litao_chi2024_ivie}. Below, we describe the updates to the two techniques.


\subsection{\tlead (V2)}
Instead of using multiple-choice questions, we updated the technique to ask short-answer questions about the functionality of the next step, requiring users to provide natural language descriptions. Each guiding question is now preceded by extensive context, to properly guide the user to think about the next step. Additionally, users receive feedback via an LLM prompt that checks their response, offering hints or pointing out missing detail. We also added on-hover explanations for each line of revealed code.

We improved question quality by enhancing the LLM prompt with chain-of-thought prompting techniques \cite{wei2022chain_of_thought}. The prompt first decomposes the code solution into a hierarchical JSON, explaining decisions at the subgoal level to each line of code. We found that this additional step significantly improved the quality of the guiding questions.

\subsection{\ttrace (V2)}
Previously, value prediction questions focused on single lines of code, which participants found too easy. The technique now asks about blocks of code, such as loops, conditionals, or multiple lines with similar goal. Additionally, users may need to predict up to two variables to address more complex scenarios. The technique is also updated to include on-hover explanations for each line.

Additionally, we merged this technique with the technique from \texplain. After predicting variable values, users are prompted to explain the purpose of the highlighted  code. This was intended to match the new difficulty level of \tlead (V2), which now requires explanations for each line rather than multiple-choice responses.

\section{Study 2: Pre vs. Post Engagement}\label{sec:second_study}
Building upon Study 1, we designed Study 2 to provide a deeper examination of the differences between the two updated techniques. While Study 1 had participants engage with only one technique, Study 2 adopted a within-subjects design to allow for direct comparison across techniques. Our objective was to evaluate:
\begin{itemize}
    \item \textbf{RQ1 [Performance]:} How effectively do the updated cognitive engagement techniques affect the transfer of learned programming concepts to isomorphic coding tasks without AI assistance?
    \item \textbf{RQ2 [Friction]:} What level of friction do the updated techniques introduce, as measured by task completion time and perceived cognitive load?
    \item \textbf{RQ3 [Metacognitive Self-Assessment]:} How do the techniques impact participants' ability to accurately assess their own performance compared to actual task performance?
    \item \textbf{RQ4 [Perceptions]:} How do participants perceive the usability and effectiveness of each technique, and how willing they are to adopt these techniques in future programming tasks?
\end{itemize}

\subsection{Methodology}
This study used the same experiment tool used in the previous study (Section \ref{sec:experiment_tool}). A key challenge in designing this study was to increase the sample size in order to detect a potential effect, based on the power analysis conducted after Study 1 (Section \ref{sec:study1_rq1_performance}). The within-subjects design would allow us to collect data on each technique from every participant. However, managing the study's duration posed another challenge. We aimed to ensure that participants had sufficient training and exposure to each technique to observe an effect, while minimizing learning effects and preventing fatigue. The following sections describe how these challenges were addressed. 

\begin{figure*}
    \centering
    \includegraphics[width=1\linewidth]{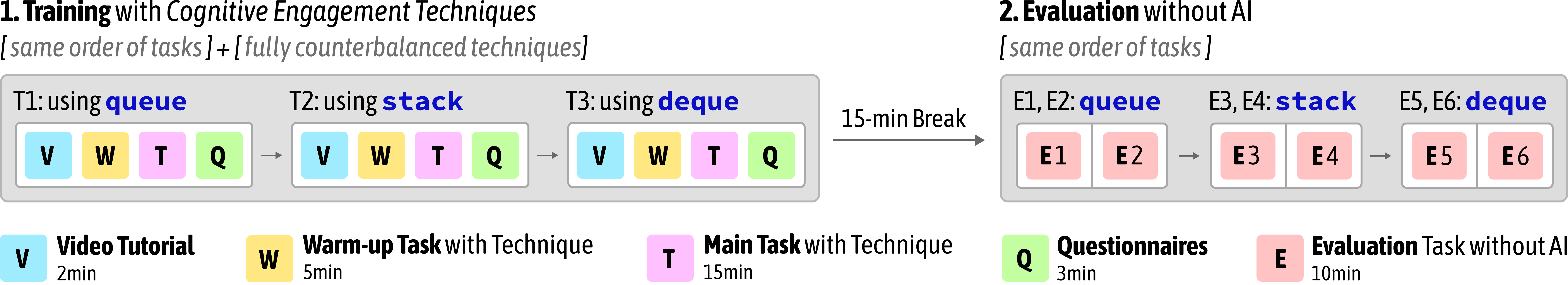}
    \caption{\textbf{Study 2 Procedure Overview}: Participants used each technique to learn complex coding tasks about three topics: queue, stack, and double-ended queue (deque). Each topic involved four steps: a short video tutorial (2 min), a warm-up task (5 min), a main task (15 min), and a questionnaire (3 min). After a 15-minute break, participants complete two evaluation tasks (10 min each) for each technique without AI assistance. Techniques during training were fully counterbalanced, while the topics had the same order in both training and evaluation phases of the experiment.}
    \label{fig:study2_procedure}
\end{figure*}


\subsubsection{Participants}
We recruited 42 participants (27 men, 15 women, 0 non-binary) from a similar population of Study 1 (students taking a data structures class), but at a different semester. Participants were pre-screened to ensure they were comfortable reading and writing in English, confident in their python programming skills, and eager to learn how to design complex algorithms. Informed consent was obtained before the study began, and all participants received \$50.

\subsection{Study Procedure}

As shown in Fig.~\ref{fig:study2_procedure}, this study consisted of two phases: training with cognitive engagement techniques and evaluation, which we explain in detail below.
\begin{itemize}
    \item \textbf{Phase 1. [Training] 3 Tasks with \tlead, \ttrace, and \tbaseline.} Unlike Study 1, this study did not include a pre-test as all participants experienced using the updated version of all three techniques: \tlead, \ttrace, and \tbaseline. The study lasted 2.5 hours, including a 15-minute break, and was conducted remotely, using MS Teams. All participants were instructed to join from a quiet and comfortable place, and share their screens to the experimenter for moderation and resolving any potential issues.
    
    Similar to the initial study, the experiment tool provided the same self-paced experience for all participants. The process started with watching a video that explained the study procedure. The training tasks with the three cognitive engagement techniques were grouped into three topics: stacks, queues, and double-ended queues (after careful consideration with the course instructor). Each topic included two tasks, a warm-up task and a training task. Topics were presented at the same order, while the techniques for which they used each topic were fully counter-balanced between the participants to reduce order effect. The assigned order of techniques for each participant was included in each participant's account information. For each topic, participants first watched a video that introduced it to the technique for which they were assigned to, followed by working on a warm-up task for 5 minutes, and then the main task of that topic for 15 minutes. Similar to the first study, before each coding task participants were given the task description and test-cases and were asked to check how confident they were in understanding the task before proceeding to working on the task with the AI and the assigned technique.
    
    After using the cognitive engagement technique to work on the task, and after finishing the main training task for each topic, participants were given the NASA Task Load Index (TLX) questionnaire, followed a 5-point Likert-question that asked how confident they are in their ability to independently write, modify, or extend code of similar complexity of the task that they just worked on without AI. This value was later used to calculate the correlation between their perceived and actual performance. Participants were then asked to take a short break, before proceeding to working on the evaluation tasks.

    \item \textbf{Phase 2. [Evaluation] 3 Topics $\times$ 2 Coding Tasks + Survey.} The evaluation tasks included three topics, each with two coding tasks. Tasks within each topic were designed to require applying the concepts used in the AI-generated code solution of their corresponding training tasks. To ensure ecological validity and simulate a task that would happen in reality, the tasks were designed to require participants to manually extend code of similar complexity to the tasks they were trained on with AI. Therefore, we designed six fill-in-the-blank tasks, in which a starter code was provided, and three to five important lines of code within each task was erased and were asked by participants to solve manually, without AI assistance. A timer was displayed on top of each evaluation task and participants were asked to skip the task after 10 minutes.
    
    Lastly, participants were given a short questionnaire in which they were asked several long-answer questions to explain their likes, dislikes, and willingness to use each of the two final code engagement techniques.

\end{itemize}

\subsubsection{Data Analysis}
To evaluate the performance on the manual coding tasks, we used a simple rubric that focused on the missing lines that participants were asked to fill. Participants only received a full mark for that line if their code for that line was fully correct. Sometimes participants' solution included additional lines of code, in such cases, they would receive full marks only if it did not change the main concept and how the task was using the given data structure for solving the task, otherwise, they did not receive marks.

In this study, we introduced another metric to determine friction created by each technique: participants' accuracy on answering the questions when participants were using each of the \tlead and \ttrace techniques which included several questions. We collected the number of questions that was asked, and how many attempts they required to correctly solve it. Participants received 100\% for answering correctly on the first attempt, 50\% if they answered correctly on the second attempt, and 0\% if they required a third attempt (as the correct answer was provided afterward). For each participant, we calculated the average accuracy for each question type, by determining the mean of their performance across all prompted questions. We looked at all three comparisons, therefore, an alpha value of $\alpha = \frac{0.05}{3} = 0.017$ is used to determine statistical significance.

\subsection{RQ1 [Performance] How effectivel are the techniques in supporting learning?}
A Friedman test did not show a statistical difference between learners' performance on the post-test evaluation tasks ($\chi^2=3.79$, $p=.150$). The \ttrace technique achieved the highest mean ($M=64.3$, $SD=34.1$), followed by \tbaseline ($M=57.6$, $SD=39.2$), and \tlead ($M=55.0$, $SD=41.1$).

\subsection{RQ2 [Friction] What level of friction do the techniques introduce?}
A Friedman test revealed a statistically significant difference in completion times among the techniques ($\chi^2(2, N=30)=37.80$, $p<.001$; completion times were not normally distributed as indicated by a Shapiro-Wilk test). The \tbaseline technique resulted in the shortest completion time ($M=470.9$, $SD=318.2$), followed by \tlead ($M=857.4$, $SD=298.6$), while \ttrace ($M=1251.2$, $SD=469.8$) took almost three times longer than the \tbaseline to complete.

In terms of the accuracy of learners' responses to the questions prompted by each technique, for the \ttrace technique, learners were asked an average of 8.3 tracing questions (to predict values) and 3.7 short-answer questions to explain the highlighted part of the code. For the \tlead technique, learners were asked an average of 8.7 questions. Learners performed significantly better on the \ttrace technique ($M=76.3\%$, $SD=14.2\%$) compared to their performance on the \tlead technique ($M=58.0\%$, $SD=12.4\%$), as indicated by a paired t-test ($p<.001$).

The Task Load Index asked participants specifically about their cognitive load to \textit{understand the AI-generated code}. For analysis, we averaged the results on the six subscales across the three techniques. A Repeated-measures ANOVA test (since task load index passed normality with a Shapiro-Wilk test) showed that task load across was statistically different across the three techniques ($F=7.4$, $p=.001$), with the \ttrace having the highest task load index ($M=67.1$, $SD=13.4$), while the task load showed no statistical difference between the \tbaseline ($M=58.9$, $SD=14.7$) and \tlead ($M=59.6$, $SD=11.9$) techniques.

On particular task load subscales, a Friedman test showed statistically significance on the temporal subscale ($p<0.001$) and a Wilcoxon signed rank showed the \ttrace technique having the highest temporal demand ($M=75.2$, $SD=19.2$). while it was similar between \tbaseline ($M=55.2$, $SD=24.1$) and the \tlead technique ($M=62.9$, $18.0$). A similar effect was seen across the physical ($p=0.019$), frustration ($p=0.021$), and performance ($p=0.039$) subscales in which the \ttrace had a higher demand, while the \tlead had a similar demand compared to the \tbaseline technique, while mental demand, perceived effort, and physical demand was similar across the three techniques.

\subsection{RQ3 [Metacognitive Self-Assessment] How do the techniques impact the ability to assess one's own performance?}
A Spearman correlation analysis examined the alignment between participants' perceived ability (self-reported via Likert-scale responses) and their actual ability (measured through post-test manual coding tasks) across the three techniques.
The \tlead intervention showed a moderate, significant correlation, $r=.26$, $p=.017$, indicating a better alignment of perceived and actual abilities. The \tbaseline technique had a weak, non-significant correlation $r=.16$, $p=.136$, and the \ttrace technique similarly showed no significant correlation, $r=.15$, $p=.171$. These results suggest that the \tlead technique supported participants' self-awareness during their engagement with the AI-generated code more effectively compared to the other techniques. 

\subsection{RQ4 [Perceptions] How do participants perceive each technique?}
After the second study, participants were asked about their likes, dislikes, and willingness to use each tool. Our qualitative analysis on their responses revealed the following themes:

\subsubsection{\tlead (V2)}
Participants frequently mentioned that the tool facilitated guided problem-solving and provided step-by-step guidance (n=26), which promoted logical thinking (n=17). They appreciated how it enhanced their understanding of the code structure (n=10), offered hints (n=11), and provided detailed feedback on mistakes (n=6). However, participants also pointed out several drawbacks: lack of flexibility in the implementation (n=10), being a time-consuming process (n=9), and difficulty in interpreting certain feedback (n=8).

\subsubsection{\ttrace (V2)}
The most common positive feedback centered around real-time variable tracking (n=18). Participants also appreciated the newly added short-answer questions and provided feedback (n=14), which helped them better understand the generated code (n=14). Some compared the experience to using debugging tools (n=10), as it enabled them to visualize code execution (n=7).

On the downside, 15 participants expressed reluctance to use the technique, citing its time-consuming nature and inefficiency for simple or long codes (n=7). Others found the questions too simple or repetitive (n=7). Additionally, some participants (n=6) reported experiencing disruptions to their workflow (n=6).

\subsection{Summary of Results}
In summary, there was no significant difference in performance across the three techniques. However, \tlead improved participants' alignment between perceived and actual ability without increasing perceived task load, despite taking 1.82x longer than \tbaseline. \ttrace, while taking 2.66x longer, introduced higher friction. Participants preferred \tlead for its support in problem-solving and computational thinking, addressing key challenges in AI-assisted learning by actively engaging users in the process. While this warrants future studies to explore the long-term impact and scalability of these techniques, \textbf{our current results point to the effectiveness of \tlead in fostering self-awareness and enhancing problem-solving skills during AI-assisted programming}.

\section{Discussion and Future Work}\label{sec:discussion}
Our results indicate that the \tlead technique achieved the highest performance in Study 1 (although not statistically significant) and demonstrated the best alignment between perceived and actual learning in Study 2, without increasing cognitive load compared to \tbaseline in both studies. To facilitate further research and usage, we have made the \tlead technique and an accompanying task builder publicly available at https://lead-and-reveal.vercel.app, enabling others to create new tasks using this technique.

In the rest of this section, we discuss the limitations of our work, synthesize design implications, and propose directions for future work.

\subsection{Limitations}
Our results are limited by the small sample size in Study 1 and limited exposure of participants to each technique in Study 2. These factors may explain the lack of observed statistical effects. Additionally, both studies focused on algorithm-heavy tasks, limiting task diversity. While tasks involving unfamiliar syntax or APIs were considered (as in Yan et al.'s study \cite{litao_chi2024_ivie}), we focused on tasks that posed a greater challenge for engagement with AI-generated code beyond simple memorization. Instead our tasks require engagement from multiple angles including decomposition, pattern recognition, data representation, abstraction, and algorithm design. Although this may limit the generalizability of our results, we believe our study addresses the most critical aspects of the issue.

In our studies, we assumed AI-generated solutions were correct, although LLMs can generate incorrect code. For techniques like \ttrace and \texplain, where the code is displayed before user interaction, this assumption is less problematic since users can read and test the code for verification. However, for techniques that involve user engagement before revealing the final solution, this could reduce perceived value of the interaction if the final code is incorrect. Although this can be mitigated by improved models and techniques (e.g., Chain-of-Thought \cite{wei2022chain_of_thought} that power recent models like GPT-o1), future techniques could verify AI-generated code by running test cases, engaging users only if the tests pass. Additionally, the techniques did not allow users to ask follow-up questions from the AI. Future research could explore follow-up interactions as another metric for user engagement with AI-generated code. Participants were also assigned to specific techniques in our experiments, we envision that in real-world settings, users would have the option to choose their preferred type of friction. Future work could explore providing this agency and examine user preferences over time. These simplifications in our study designs ensured consistency and allowed us to focus on evaluating the effectiveness of the types of friction introduced with each technique.


\subsection{Friction-Induced AI}
To ensure long-term productivity gains, researchers in human-AI interaction have proposed various concepts, such as promoting `metacognitive' reflection \cite{tankelevitch2024metacognitive_demands}, antagonistic or sycophantic AI \cite{cai2024antagonisticai}, or making AI act as coach \cite{hofman2023sports_ai}, or AI as a provocateur \cite{sarkar2024aiprovocateur}. A common theme among these concepts, is a call for action that AI should augment and not automate. In alignment with these approaches, we introduce \textit{Friction-Induced AI}, wherein the AI does not allow users to use its generated solution immediately. Instead, it engages the user in the process of generation, while challenging them (similar to \tlead) and providing opportunities for reflection. We believe this concept generalizes beyond just AI-assisted programming but human-AI interaction more generally. The term ``friction'' here does not refer to creating frustration or unnecessary difficulty. Rather, it refers to temporarily slowing the user's interaction, ensuring they engage critically, rather than passively using AI's output. However, our results---specifically in code generation---show that for effective outcomes, friction must be properly designed. Below we highlight key dimensions in effectively designing \textit{Friction-Induced AI}, particularly in AI-assisted programming.

\subsubsection{Effective Friction Type}
An interesting finding from the second study (Section \ref{sec:second_study}) was that participants performed significantly less accurately on the leading short-answer questions in \tlead (58\%) compared to \ttrace (76\%), indicating that users were more challenged by the questions in \tlead. Notably, despite the increased challenge, the use of \tlead did not result in a higher cognitive load compared to \tbaseline, whereas \ttrace did. This suggests that effective and beneficial friction should be designed to engage users in meaningful activities, similar to \tlead, which engaged users in the problem-solving aspect of the task. The technique also allowed reflection in a short, interactive conversation with the AI. Similarly, \tparsons provided a scaffolded approach to constructing code from scrambled code blocks, which previous research has shown to be both fun and engaging without imposing high cognitive demands \cite{denny2008parsons, parsons2006parsons, weinman2021faded_parsons}. 

Additionally, the \tlead technique inherently has gamification in its engagement and learning process: the next line of code is revealed only after the user successfully explains the action required in that line. This could be a rewarding experience and increase motivation while engaging the user in a process that mirrors real-world problem-solving. However, not all tasks are solved linearly, from top to bottom. Often, code is constructed from a combination of patterns that are distributed throughout the code. Future tools should therefore focus on decomposing code hierarchically and guiding users in the hierarchical decision making process.

In contrast, effective friction should avoid engaging the user in simple and repetitive tasks like typing over generated code (\tguidedwriteover), or merely translating pseudocode to code (\tpseudo). It should also avoid tasks where the engagement does not create new knowledge like repeatedly tracing values of different variables (\ttrace), or confusing the user by mixing correct and incorrect code (\tverify).


\subsubsection{Beyond Code Generation}
AI-assisted programming extends beyond mere code generation or completion. It includes various scenarios where introducing appropriate friction can enhance cognitive engagement and help prevent skill degradation. In these contexts, AI operates autonomously, solving a user's problem like an agent without additional user involvement. Here we discuss two particular scenarios: debugging and AI-assisted decision making.

First, when AI is used for debugging and fixing a user's code---particularly for novice or end-user programmers \cite{ko2004learning_barriers_eup}---it often generates the fixed solution. However, debugging is a fundamental computational thinking skill \cite{wing2006computational_thinking}, and must be practiced to ensure programming proficiency. Researchers have already explored introducing similar forms of friction to support help-seeking behaviors. CodeAid \cite{kazemitabaar_chi24_codeaid} highlights incorrect lines of code with suggested fixes that users must apply themselves. Similarly, CodeTailor \cite{hou2024codetailor} generates a scrambled version of the fixed code, requiring users to rearrange it, similar to \tparsons. Another form of meaningful friction could involve engaging the user in the step-by-step debugging process alongside the AI, similar to \tlead. This approach could enhance debugging skills while offering a greater sense of accomplishment.

Second, when AI is being used to make decisions for programming tasks---especially in scenarios involving complex trade-offs, such as machine learning, data analysis \cite{gu2024analysts_ai_assistance, gu2024verify_ai_assistance}, and software design \cite{white2024ai_software_design}---it is crucial to encourage programmers to think critically about these decisions. Reflecting on these decisions promotes their critical thinking and problem-solving skills, and their ability to verify AI-generated code and decisions. Researchers are already developing tools to assist novices in making better software design and architectural decisions with generative AI \cite{pace2024novice_architects_genai}. However, when AI autonomously solves tasks for the user, there is a risk that this could negatively impact their decision-making skills. A more engaged, user-involved AI assistance might result in better decisions and improved mental models, an area that future research should explore.

\subsubsection{Beyond Educational Contexts}
Lastly, an important design implication emerges from the synergy between adding friction and enhanced verification in AI-assisted programming. Although we studied friction in the context of programming with novices learning about complex algorithmic tasks, but friction can potentially be suitable in productivity scenarios as well. Friction-induced AI can potentially prevent long-term productivity losses, as well as enhance short-term verification of AI outputs, leading to immediate productivity gains. Introducing friction, like cognitive forcing functions, has been shown to support AI-assisted decision-making and reduce over-reliance on AI \cite{buccinca2021trust}. Similarly, adding intervention points and employing progressive disclosure has been another type of friction that has been demonstrated to increase programmers' sense of control and ability to verify outputs during AI-assisted data analysis \cite{kazemitabaar_uist24_steering_verification}.

This dual benefit underscores how adding meaningful and effective friction could make it suitable for productivity scenarios and not just for educational contexts. It not only improves short-term verification and productivity but also helps prevent long-term productivity loss by augmenting human cognition and encouraging higher-order cognitive engagement. As a result, techniques like \tlead could be designed to incorporate intervention points that allows users to progressively verify and correct the AI's output, while also promoting deeper learning as code is generated.

\section{Conclusion}

Over-reliance on AI poses considerable risk to the development of computational thinking skills among novice programmers.
However, as generative AI continues to change the landscape of programming, it is essential to harness its benefits by understanding how to effectively integrate it into educational settings.
In this paper, we introduced the concept of \emph{friction-induced AI}, which requires users to cognitively engage with AI-generated content (specifically code), to enhance short-term productivity gains while preventing long-term productivity loss due to over-reliance on AI.

We systematically explored the design space of cognitive engagement techniques, developing seven distinct interfaces that introduce varying levels and types of engagement.
Through an iterative design process and two empirical evaluations, we identified that the \tlead technique balances imposed friction, learning gains, and improved self-assessment accuracy.
Specifically, \tlead improved the alignment between learns' perceived and actual coding abilities without increasing cognitive load, demonstrating its potential to improve learning and prevent over-reliance.
Beneficial friction can be viewed as a supportive scaffold that empowers students rather than acting as an obstacle.
By requiring users to engage actively, these techniques promote metacognitive reflection and improve problem-solving skills.
Future research should explore more diverse programming tasks and user-driven engagement strategies to further validate these findings. 
Ultimately, designing AI tools that support deeper cognitive reflection remains crucial to maintaining both short-term productivity and long-term skill development in programming.

\begin{acks}
We would like to thank the computing education researchers for their valuable feedback on our techniques. We also sincerely appreciate the course instructors for their input on the design of the programming tasks, and we are grateful to the students who participated in our experiments.
\end{acks}

\bibliographystyle{ACM-Reference-Format}
\bibliography{references}


\begin{thebibliography}{123}


\ifx \showCODEN    \undefined \def \showCODEN     #1{\unskip}     \fi
\ifx \showDOI      \undefined \def \showDOI       #1{#1}\fi
\ifx \showISBNx    \undefined \def \showISBNx     #1{\unskip}     \fi
\ifx \showISBNxiii \undefined \def \showISBNxiii  #1{\unskip}     \fi
\ifx \showISSN     \undefined \def \showISSN      #1{\unskip}     \fi
\ifx \showLCCN     \undefined \def \showLCCN      #1{\unskip}     \fi
\ifx \shownote     \undefined \def \shownote      #1{#1}          \fi
\ifx \showarticletitle \undefined \def \showarticletitle #1{#1}   \fi
\ifx \showURL      \undefined \def \showURL       {\relax}        \fi
\providecommand\bibfield[2]{#2}
\providecommand\bibinfo[2]{#2}
\providecommand\natexlab[1]{#1}
\providecommand\showeprint[2][]{arXiv:#2}

\bibitem[Altadmri and Brown(2015)]%
        {Altadmri2015SIGCSE}
\bibfield{author}{\bibinfo{person}{Amjad Altadmri} {and} \bibinfo{person}{Neil~C.C. Brown}.} \bibinfo{year}{2015}\natexlab{}.
\newblock \showarticletitle{37 Million Compilations: Investigating Novice Programming Mistakes in Large-Scale Student Data}. In \bibinfo{booktitle}{\emph{Proceedings of the 46th ACM Technical Symposium on Computer Science Education}} (Kansas City, Missouri, USA) \emph{(\bibinfo{series}{SIGCSE '15})}. \bibinfo{publisher}{ACM}, \bibinfo{address}{New York, NY, USA}, \bibinfo{pages}{522--527}.
\newblock
\showISBNx{978-1-4503-2966-8}
\urldef\tempurl%
\url{https://doi.org/10.1145/2676723.2677258}
\showDOI{\tempurl}


\bibitem[Bastani et~al\mbox{.}(2024)]%
        {bastani2024genAi_harm_learning}
\bibfield{author}{\bibinfo{person}{Hamsa Bastani}, \bibinfo{person}{Osbert Bastani}, \bibinfo{person}{Alp Sungu}, \bibinfo{person}{Haosen Ge}, \bibinfo{person}{Özge Kabakcı}, {and} \bibinfo{person}{Rei Mariman}.} \bibinfo{year}{2024}\natexlab{}.
\newblock \bibinfo{booktitle}{\emph{Generative AI Can Harm Learning}}.
\newblock \bibinfo{type}{Research Paper}. \bibinfo{institution}{The Wharton School}.
\newblock
\urldef\tempurl%
\url{https://doi.org/10.2139/ssrn.4895486}
\showDOI{\tempurl}
\newblock
\shownote{Available at SSRN: https://ssrn.com/abstract=4895486 or http://dx.doi.org/10.2139/ssrn.4895486}.


\bibitem[Becker(2016a)]%
        {Becker2016SIGCSE}
\bibfield{author}{\bibinfo{person}{Brett~A. Becker}.} \bibinfo{year}{2016}\natexlab{a}.
\newblock \showarticletitle{An Effective Approach to Enhancing Compiler Error Messages}. In \bibinfo{booktitle}{\emph{Proceedings of the 47th ACM Technical Symposium on Computing Science Education}} (Memphis, Tennessee, USA) \emph{(\bibinfo{series}{SIGCSE '16})}. \bibinfo{publisher}{ACM}, \bibinfo{address}{New York, NY, USA}, \bibinfo{pages}{126--131}.
\newblock
\showISBNx{978-1-4503-3685-7}
\urldef\tempurl%
\url{https://doi.org/10.1145/2839509.2844584}
\showDOI{\tempurl}


\bibitem[Becker(2016b)]%
        {Becker2016ITiCSE}
\bibfield{author}{\bibinfo{person}{Brett~A. Becker}.} \bibinfo{year}{2016}\natexlab{b}.
\newblock \showarticletitle{A New Metric to Quantify Repeated Compiler Errors for Novice Programmers}. In \bibinfo{booktitle}{\emph{Proceedings of the 2016 ACM Conference on Innovation and Technology in Computer Science Education}} (Arequipa, Peru) \emph{(\bibinfo{series}{ITiCSE '16})}. \bibinfo{publisher}{ACM}, \bibinfo{address}{New York, NY, USA}, \bibinfo{pages}{296--301}.
\newblock
\showISBNx{978-1-4503-4231-5}
\urldef\tempurl%
\url{https://doi.org/10.1145/2899415.2899463}
\showDOI{\tempurl}


\bibitem[Becker et~al\mbox{.}(2023)]%
        {Becker2023CodingWasHard}
\bibfield{author}{\bibinfo{person}{Brett~A. Becker}, \bibinfo{person}{Paul Denny}, \bibinfo{person}{James Finnie-Ansley}, \bibinfo{person}{Andrew Luxton-Reilly}, \bibinfo{person}{James Prather}, {and} \bibinfo{person}{Eddie~Antonio Santos}.} \bibinfo{year}{2023}\natexlab{}.
\newblock \showarticletitle{Programming Is Hard - Or at Least It Used to Be: Educational Opportunities and Challenges of AI Code Generation}. In \bibinfo{booktitle}{\emph{Proceedings of the 54th ACM Technical Symposium on Computer Science Education V. 1}} (Toronto ON, Canada) \emph{(\bibinfo{series}{SIGCSE 2023})}. \bibinfo{publisher}{Association for Computing Machinery}, \bibinfo{address}{New York, NY, USA}, \bibinfo{pages}{500–506}.
\newblock
\showISBNx{9781450394314}
\urldef\tempurl%
\url{https://doi.org/10.1145/3545945.3569759}
\showDOI{\tempurl}


\bibitem[Bell et~al\mbox{.}(1998)]%
        {bell1998csunplugged}
\bibfield{author}{\bibinfo{person}{Timothy~C Bell}, \bibinfo{person}{Ian~H Witten}, {and} \bibinfo{person}{Mike Fellows}.} \bibinfo{year}{1998}\natexlab{}.
\newblock \bibinfo{booktitle}{\emph{Computer Science Unplugged: Off-line Activities and Games for All Ages} (\bibinfo{edition}{1st} ed.)}.
\newblock \bibinfo{publisher}{Computer Science Unplugged}.
\newblock


\bibitem[Bloom et~al\mbox{.}(1956)]%
        {bloom1956taxonomy}
\bibfield{author}{\bibinfo{person}{Benjamin~S Bloom}, \bibinfo{person}{Max~D Engelhart}, \bibinfo{person}{Edward~J Furst}, \bibinfo{person}{Walker~H Hill}, \bibinfo{person}{David~R Krathwohl}, {et~al\mbox{.}}} \bibinfo{year}{1956}\natexlab{}.
\newblock \bibinfo{booktitle}{\emph{Taxonomy of Educational Objectives: The Classification of Educational Goals. Handbook 1: Cognitive Domain}}.
\newblock \bibinfo{publisher}{Longman New York}.
\newblock


\bibitem[Bonar and Soloway(2013)]%
        {bonar2013preprogramming}
\bibfield{author}{\bibinfo{person}{Jeffrey Bonar} {and} \bibinfo{person}{Elliot Soloway}.} \bibinfo{year}{2013}\natexlab{}.
\newblock \showarticletitle{Preprogramming knowledge: A major source of misconceptions in novice programmers}.
\newblock In \bibinfo{booktitle}{\emph{Studying the novice programmer}}. \bibinfo{publisher}{Psychology Press}, \bibinfo{pages}{325--353}.
\newblock


\bibitem[Brackmann et~al\mbox{.}(2017)]%
        {brackmann2017csunplugged}
\bibfield{author}{\bibinfo{person}{Christian~P. Brackmann}, \bibinfo{person}{Marcos Rom\'{a}n-Gonz\'{a}lez}, \bibinfo{person}{Gregorio Robles}, \bibinfo{person}{Jes\'{u}s Moreno-Le\'{o}n}, \bibinfo{person}{Ana Casali}, {and} \bibinfo{person}{Dante Barone}.} \bibinfo{year}{2017}\natexlab{}.
\newblock \showarticletitle{Development of Computational Thinking Skills through Unplugged Activities in Primary School}. In \bibinfo{booktitle}{\emph{Proceedings of the 12th Workshop on Primary and Secondary Computing Education}} (Nijmegen, Netherlands) \emph{(\bibinfo{series}{WiPSCE '17})}. \bibinfo{publisher}{Association for Computing Machinery}, \bibinfo{address}{New York, NY, USA}, \bibinfo{pages}{65–72}.
\newblock
\showISBNx{9781450354288}
\urldef\tempurl%
\url{https://doi.org/10.1145/3137065.3137069}
\showDOI{\tempurl}


\bibitem[Brandt et~al\mbox{.}(2009)]%
        {brandt2009opportunistic}
\bibfield{author}{\bibinfo{person}{Joel Brandt}, \bibinfo{person}{Philip~J. Guo}, \bibinfo{person}{Joel Lewenstein}, \bibinfo{person}{Mira Dontcheva}, {and} \bibinfo{person}{Scott~R. Klemmer}.} \bibinfo{year}{2009}\natexlab{}.
\newblock \showarticletitle{Two studies of opportunistic programming: interleaving web foraging, learning, and writing code}. In \bibinfo{booktitle}{\emph{Proceedings of the SIGCHI Conference on Human Factors in Computing Systems}} (Boston, MA, USA) \emph{(\bibinfo{series}{CHI '09})}. \bibinfo{publisher}{Association for Computing Machinery}, \bibinfo{address}{New York, NY, USA}, \bibinfo{pages}{1589–1598}.
\newblock
\showISBNx{9781605582467}
\urldef\tempurl%
\url{https://doi.org/10.1145/1518701.1518944}
\showDOI{\tempurl}


\bibitem[Brown and Altadmri(2014)]%
        {Brown2014ICER}
\bibfield{author}{\bibinfo{person}{Neil~C.C. Brown} {and} \bibinfo{person}{Amjad Altadmri}.} \bibinfo{year}{2014}\natexlab{}.
\newblock \showarticletitle{Investigating Novice Programming Mistakes: Educator Beliefs vs. Student Data}. In \bibinfo{booktitle}{\emph{Proceedings of the Tenth Annual Conference on International Computing Education Research}} (Glasgow, Scotland, United Kingdom) \emph{(\bibinfo{series}{ICER '14})}. \bibinfo{publisher}{ACM}, \bibinfo{address}{New York, NY, USA}, \bibinfo{pages}{43--50}.
\newblock
\showISBNx{978-1-4503-2755-8}
\urldef\tempurl%
\url{https://doi.org/10.1145/2632320.2632343}
\showDOI{\tempurl}


\bibitem[Bu\c{c}inca et~al\mbox{.}(2021a)]%
        {buccinca2021trust}
\bibfield{author}{\bibinfo{person}{Zana Bu\c{c}inca}, \bibinfo{person}{Maja~Barbara Malaya}, {and} \bibinfo{person}{Krzysztof~Z. Gajos}.} \bibinfo{year}{2021}\natexlab{a}.
\newblock \showarticletitle{To Trust or to Think: Cognitive Forcing Functions Can Reduce Overreliance on AI in AI-assisted Decision-making}.
\newblock \bibinfo{journal}{\emph{Proc. ACM Hum.-Comput. Interact.}} \bibinfo{volume}{5}, \bibinfo{number}{CSCW1}, Article \bibinfo{articleno}{188} (\bibinfo{date}{apr} \bibinfo{year}{2021}), \bibinfo{numpages}{21}~pages.
\newblock
\urldef\tempurl%
\url{https://doi.org/10.1145/3449287}
\showDOI{\tempurl}


\bibitem[Bu\c{c}inca et~al\mbox{.}(2021b)]%
        {buccinca_cscw21_overreliance_ai}
\bibfield{author}{\bibinfo{person}{Zana Bu\c{c}inca}, \bibinfo{person}{Maja~Barbara Malaya}, {and} \bibinfo{person}{Krzysztof~Z. Gajos}.} \bibinfo{year}{2021}\natexlab{b}.
\newblock \showarticletitle{To Trust or to Think: Cognitive Forcing Functions Can Reduce Overreliance on AI in AI-assisted Decision-making}.
\newblock \bibinfo{journal}{\emph{Proc. ACM Hum.-Comput. Interact.}} \bibinfo{volume}{5}, \bibinfo{number}{CSCW1} (\bibinfo{date}{apr} \bibinfo{year}{2021}).
\newblock


\bibitem[Cai et~al\mbox{.}(2024)]%
        {cai2024antagonisticai}
\bibfield{author}{\bibinfo{person}{Alice Cai}, \bibinfo{person}{Ian Arawjo}, {and} \bibinfo{person}{Elena~L. Glassman}.} \bibinfo{year}{2024}\natexlab{}.
\newblock \bibinfo{title}{Antagonistic AI}.
\newblock
\newblock
\showeprint[arxiv]{2402.07350}
\urldef\tempurl%
\url{https://arxiv.org/abs/2402.07350}
\showURL{%
\tempurl}


\bibitem[Chatterjee et~al\mbox{.}(2020)]%
        {chatterjee2020novice_se_stackoverflow}
\bibfield{author}{\bibinfo{person}{Preetha Chatterjee}, \bibinfo{person}{Minji Kong}, {and} \bibinfo{person}{Lori Pollock}.} \bibinfo{year}{2020}\natexlab{}.
\newblock \showarticletitle{Finding help with programming errors: An exploratory study of novice software engineers' focus in stack overflow posts}.
\newblock \bibinfo{journal}{\emph{Journal of Systems and Software}}  \bibinfo{volume}{159} (\bibinfo{year}{2020}), \bibinfo{pages}{110454}.
\newblock
\showISSN{0164-1212}
\urldef\tempurl%
\url{https://doi.org/10.1016/j.jss.2019.110454}
\showDOI{\tempurl}


\bibitem[Chi and Wylie(2014)]%
        {chi2014icap_framework}
\bibfield{author}{\bibinfo{person}{Michelene T.~H. Chi} {and} \bibinfo{person}{Ruth Wylie}.} \bibinfo{year}{2014}\natexlab{}.
\newblock \showarticletitle{The ICAP Framework: Linking Cognitive Engagement to Active Learning Outcomes}.
\newblock \bibinfo{journal}{\emph{Educational Psychologist}} \bibinfo{volume}{49}, \bibinfo{number}{4} (\bibinfo{year}{2014}), \bibinfo{pages}{219--243}.
\newblock
\urldef\tempurl%
\url{https://doi.org/10.1080/00461520.2014.965823}
\showDOI{\tempurl}


\bibitem[Denny et~al\mbox{.}(2024a)]%
        {denny2024prompt_problems}
\bibfield{author}{\bibinfo{person}{Paul Denny}, \bibinfo{person}{Juho Leinonen}, \bibinfo{person}{James Prather}, \bibinfo{person}{Andrew Luxton-Reilly}, \bibinfo{person}{Thezyrie Amarouche}, \bibinfo{person}{Brett~A. Becker}, {and} \bibinfo{person}{Brent~N. Reeves}.} \bibinfo{year}{2024}\natexlab{a}.
\newblock \showarticletitle{Prompt Problems: A New Programming Exercise for the Generative AI Era}. In \bibinfo{booktitle}{\emph{Proceedings of the 55th ACM Technical Symposium on Computer Science Education V. 1}} (Portland, OR, USA) \emph{(\bibinfo{series}{SIGCSE 2024})}. \bibinfo{publisher}{Association for Computing Machinery}, \bibinfo{address}{New York, NY, USA}, \bibinfo{pages}{296–302}.
\newblock
\showISBNx{9798400704239}
\urldef\tempurl%
\url{https://doi.org/10.1145/3626252.3630909}
\showDOI{\tempurl}


\bibitem[Denny et~al\mbox{.}(2014)]%
        {Denny2014ITiCSE}
\bibfield{author}{\bibinfo{person}{Paul Denny}, \bibinfo{person}{Andrew Luxton-Reilly}, {and} \bibinfo{person}{Dave Carpenter}.} \bibinfo{year}{2014}\natexlab{}.
\newblock \showarticletitle{Enhancing Syntax Error Messages Appears Ineffectual}. In \bibinfo{booktitle}{\emph{Proceedings of the 2014 Conference on Innovation \&\#38; Technology in Computer Science Education}} (Uppsala, Sweden) \emph{(\bibinfo{series}{ITiCSE '14})}. \bibinfo{publisher}{ACM}, \bibinfo{address}{New York, NY, USA}, \bibinfo{pages}{273--278}.
\newblock
\showISBNx{978-1-4503-2833-3}
\urldef\tempurl%
\url{https://doi.org/10.1145/2591708.2591748}
\showDOI{\tempurl}


\bibitem[Denny et~al\mbox{.}(2008)]%
        {denny2008parsons}
\bibfield{author}{\bibinfo{person}{Paul Denny}, \bibinfo{person}{Andrew Luxton-Reilly}, {and} \bibinfo{person}{Beth Simon}.} \bibinfo{year}{2008}\natexlab{}.
\newblock \showarticletitle{Evaluating a new exam question: Parsons problems}. In \bibinfo{booktitle}{\emph{Proceedings of the Fourth International Workshop on Computing Education Research}} (Sydney, Australia) \emph{(\bibinfo{series}{ICER '08})}. \bibinfo{publisher}{Association for Computing Machinery}, \bibinfo{address}{New York, NY, USA}, \bibinfo{pages}{113–124}.
\newblock
\showISBNx{9781605582160}
\urldef\tempurl%
\url{https://doi.org/10.1145/1404520.1404532}
\showDOI{\tempurl}


\bibitem[Denny et~al\mbox{.}(2024b)]%
        {denny2024explain_code_purpose}
\bibfield{author}{\bibinfo{person}{Paul Denny}, \bibinfo{person}{David~H. Smith}, \bibinfo{person}{Max Fowler}, \bibinfo{person}{James Prather}, \bibinfo{person}{Brett~A. Becker}, {and} \bibinfo{person}{Juho Leinonen}.} \bibinfo{year}{2024}\natexlab{b}.
\newblock \showarticletitle{Explaining Code with a Purpose: An Integrated Approach for Developing Code Comprehension and Prompting Skills}. In \bibinfo{booktitle}{\emph{Proceedings of the 2024 on Innovation and Technology in Computer Science Education V. 1}} (Milan, Italy) \emph{(\bibinfo{series}{ITiCSE 2024})}. \bibinfo{publisher}{Association for Computing Machinery}, \bibinfo{address}{New York, NY, USA}, \bibinfo{pages}{283–289}.
\newblock
\showISBNx{9798400706004}
\urldef\tempurl%
\url{https://doi.org/10.1145/3649217.3653587}
\showDOI{\tempurl}


\bibitem[D{\'\i}az-Pace et~al\mbox{.}(2024)]%
        {pace2024novice_architects_genai}
\bibfield{author}{\bibinfo{person}{J~Andr{\'e}s D{\'\i}az-Pace}, \bibinfo{person}{Antonela Tommasel}, {and} \bibinfo{person}{Rafael Capilla}.} \bibinfo{year}{2024}\natexlab{}.
\newblock \showarticletitle{Helping Novice Architects to Make Quality Design Decisions Using an LLM-Based Assistant}. In \bibinfo{booktitle}{\emph{Software Architecture}}. \bibinfo{publisher}{Springer Nature Switzerland}, \bibinfo{address}{Cham}, \bibinfo{pages}{324--332}.
\newblock
\showISBNx{978-3-031-70797-1}


\bibitem[Du~Boulay(2013)]%
        {du2013some}
\bibfield{author}{\bibinfo{person}{Benedict Du~Boulay}.} \bibinfo{year}{2013}\natexlab{}.
\newblock \showarticletitle{Some difficulties of learning to program}.
\newblock In \bibinfo{booktitle}{\emph{Studying the novice programmer}}. \bibinfo{publisher}{Psychology Press}, \bibinfo{pages}{283--299}.
\newblock


\bibitem[Ebrahimi(1994)]%
        {Ebrahimi1994IJHCS}
\bibfield{author}{\bibinfo{person}{Alireza Ebrahimi}.} \bibinfo{year}{1994}\natexlab{}.
\newblock \showarticletitle{Novice programmer errors: Language constructs and plan composition}.
\newblock \bibinfo{journal}{\emph{International Journal of Human Computer Studies}} \bibinfo{volume}{41}, \bibinfo{number}{4} (\bibinfo{year}{1994}), \bibinfo{pages}{457--480}.
\newblock


\bibitem[Ericson et~al\mbox{.}(2018)]%
        {ericson2018adaptive_parsons}
\bibfield{author}{\bibinfo{person}{Barbara~J. Ericson}, \bibinfo{person}{James~D. Foley}, {and} \bibinfo{person}{Jochen Rick}.} \bibinfo{year}{2018}\natexlab{}.
\newblock \showarticletitle{Evaluating the Efficiency and Effectiveness of Adaptive Parsons Problems}. In \bibinfo{booktitle}{\emph{Proceedings of the 2018 ACM Conference on International Computing Education Research}} (Espoo, Finland) \emph{(\bibinfo{series}{ICER '18})}. \bibinfo{publisher}{Association for Computing Machinery}, \bibinfo{address}{New York, NY, USA}, \bibinfo{pages}{60–68}.
\newblock
\showISBNx{9781450356282}
\urldef\tempurl%
\url{https://doi.org/10.1145/3230977.3231000}
\showDOI{\tempurl}


\bibitem[Ericson et~al\mbox{.}(2017)]%
        {ericson2017parsons}
\bibfield{author}{\bibinfo{person}{Barbara~J. Ericson}, \bibinfo{person}{Lauren~E. Margulieux}, {and} \bibinfo{person}{Jochen Rick}.} \bibinfo{year}{2017}\natexlab{}.
\newblock \showarticletitle{Solving parsons problems versus fixing and writing code}. In \bibinfo{booktitle}{\emph{Proceedings of the 17th Koli Calling International Conference on Computing Education Research}} (Koli, Finland) \emph{(\bibinfo{series}{Koli Calling '17})}. \bibinfo{publisher}{Association for Computing Machinery}, \bibinfo{address}{New York, NY, USA}, \bibinfo{pages}{20–29}.
\newblock
\showISBNx{9781450353014}
\urldef\tempurl%
\url{https://doi.org/10.1145/3141880.3141895}
\showDOI{\tempurl}


\bibitem[Ericsson et~al\mbox{.}(1993)]%
        {ericsson1993deliberate_practice}
\bibfield{author}{\bibinfo{person}{K~Anders Ericsson}, \bibinfo{person}{Ralf~T Krampe}, {and} \bibinfo{person}{Clemens Tesch-R{\"o}mer}.} \bibinfo{year}{1993}\natexlab{}.
\newblock \showarticletitle{The role of deliberate practice in the acquisition of expert performance.}
\newblock \bibinfo{journal}{\emph{Psychological review}} \bibinfo{volume}{100}, \bibinfo{number}{3} (\bibinfo{year}{1993}), \bibinfo{pages}{363}.
\newblock
\urldef\tempurl%
\url{https://doi.org/10.1037/0033-295X.100.3.363}
\showDOI{\tempurl}


\bibitem[Fast et~al\mbox{.}(2014)]%
        {Fast2014CHI}
\bibfield{author}{\bibinfo{person}{Ethan Fast}, \bibinfo{person}{Daniel Steffee}, \bibinfo{person}{Lucy Wang}, \bibinfo{person}{Joel~R. Brandt}, {and} \bibinfo{person}{Michael~S. Bernstein}.} \bibinfo{year}{2014}\natexlab{}.
\newblock \showarticletitle{Emergent, Crowd-scale Programming Practice in the IDE}. In \bibinfo{booktitle}{\emph{Proceedings of the SIGCHI Conference on Human Factors in Computing Systems}} (Toronto, Ontario, Canada) \emph{(\bibinfo{series}{CHI '14})}. \bibinfo{publisher}{ACM}, \bibinfo{address}{New York, NY, USA}, \bibinfo{pages}{2491--2500}.
\newblock
\showISBNx{978-1-4503-2473-1}
\urldef\tempurl%
\url{https://doi.org/10.1145/2556288.2556998}
\showDOI{\tempurl}


\bibitem[Finnie-Ansley et~al\mbox{.}(2023)]%
        {finnie2024ai_cs2_tasks}
\bibfield{author}{\bibinfo{person}{James Finnie-Ansley}, \bibinfo{person}{Paul Denny}, \bibinfo{person}{Andrew Luxton-Reilly}, \bibinfo{person}{Eddie~Antonio Santos}, \bibinfo{person}{James Prather}, {and} \bibinfo{person}{Brett~A. Becker}.} \bibinfo{year}{2023}\natexlab{}.
\newblock \showarticletitle{My AI Wants to Know if This Will Be on the Exam: Testing OpenAI’s Codex on CS2 Programming Exercises}. In \bibinfo{booktitle}{\emph{Proceedings of the 25th Australasian Computing Education Conference}} (Melbourne, VIC, Australia) \emph{(\bibinfo{series}{ACE '23})}. \bibinfo{publisher}{Association for Computing Machinery}, \bibinfo{address}{New York, NY, USA}, \bibinfo{pages}{97–104}.
\newblock
\showISBNx{9781450399418}
\urldef\tempurl%
\url{https://doi.org/10.1145/3576123.3576134}
\showDOI{\tempurl}


\bibitem[Flowers et~al\mbox{.}(2004)]%
        {Flowers2004FIE}
\bibfield{author}{\bibinfo{person}{T. Flowers}, \bibinfo{person}{J. Jackson}, {and} \bibinfo{person}{C. Carver}.} \bibinfo{year}{2004}\natexlab{}.
\newblock \showarticletitle{Empowering students and building confidence in novice programmers through Gauntlet}. In \bibinfo{booktitle}{\emph{34th Annual Frontiers in Education, 2004. FIE 2004.(FIE)}}, Vol.~\bibinfo{volume}{00}. \bibinfo{pages}{T3H/10--T3H/13 Vol. 1}.
\newblock
\urldef\tempurl%
\url{https://doi.org/10.1109/FIE.2004.1408551}
\showDOI{\tempurl}


\bibitem[Gajos and Mamykina(2022)]%
        {gajos2022incidentallearning}
\bibfield{author}{\bibinfo{person}{Krzysztof~Z. Gajos} {and} \bibinfo{person}{Lena Mamykina}.} \bibinfo{year}{2022}\natexlab{}.
\newblock \showarticletitle{Do People Engage Cognitively with AI? Impact of AI Assistance on Incidental Learning}. In \bibinfo{booktitle}{\emph{Proceedings of the 27th International Conference on Intelligent User Interfaces}} (Helsinki, Finland) \emph{(\bibinfo{series}{IUI '22})}. \bibinfo{publisher}{Association for Computing Machinery}, \bibinfo{address}{New York, NY, USA}, \bibinfo{pages}{794–806}.
\newblock
\showISBNx{9781450391443}
\urldef\tempurl%
\url{https://doi.org/10.1145/3490099.3511138}
\showDOI{\tempurl}


\bibitem[Galenson et~al\mbox{.}(2014)]%
        {Galenson2014ICSE}
\bibfield{author}{\bibinfo{person}{Joel Galenson}, \bibinfo{person}{Philip Reames}, \bibinfo{person}{Rastislav Bodik}, \bibinfo{person}{Bj\"{o}rn Hartmann}, {and} \bibinfo{person}{Koushik Sen}.} \bibinfo{year}{2014}\natexlab{}.
\newblock \showarticletitle{CodeHint: Dynamic and Interactive Synthesis of Code Snippets}. In \bibinfo{booktitle}{\emph{Proceedings of the 36th International Conference on Software Engineering}} (Hyderabad, India) \emph{(\bibinfo{series}{ICSE 2014})}. \bibinfo{publisher}{ACM}, \bibinfo{address}{New York, NY, USA}, \bibinfo{pages}{653--663}.
\newblock
\showISBNx{978-1-4503-2756-5}
\urldef\tempurl%
\url{https://doi.org/10.1145/2568225.2568250}
\showDOI{\tempurl}


\bibitem[Gaweda et~al\mbox{.}(2020)]%
        {gaweda2020typing_exercises}
\bibfield{author}{\bibinfo{person}{Adam~M. Gaweda}, \bibinfo{person}{Collin~F. Lynch}, \bibinfo{person}{Nathan Seamon}, \bibinfo{person}{Gabriel Silva~de Oliveira}, {and} \bibinfo{person}{Alay Deliwa}.} \bibinfo{year}{2020}\natexlab{}.
\newblock \showarticletitle{Typing Exercises as Interactive Worked Examples for Deliberate Practice in CS Courses}. In \bibinfo{booktitle}{\emph{Proceedings of the Twenty-Second Australasian Computing Education Conference}} (Melbourne, VIC, Australia) \emph{(\bibinfo{series}{ACE'20})}. \bibinfo{publisher}{Association for Computing Machinery}, \bibinfo{address}{New York, NY, USA}, \bibinfo{pages}{105–113}.
\newblock
\showISBNx{9781450376860}
\urldef\tempurl%
\url{https://doi.org/10.1145/3373165.3373177}
\showDOI{\tempurl}


\bibitem[Ghimire and Edwards(2024)]%
        {ghimire2024coding}
\bibfield{author}{\bibinfo{person}{Aashish Ghimire} {and} \bibinfo{person}{John Edwards}.} \bibinfo{year}{2024}\natexlab{}.
\newblock \showarticletitle{Coding with AI: How Are Tools Like ChatGPT Being Used by Students in Foundational Programming Courses}. In \bibinfo{booktitle}{\emph{Artificial Intelligence in Education}}, \bibfield{editor}{\bibinfo{person}{Andrew~M. Olney}, \bibinfo{person}{Irene-Angelica Chounta}, \bibinfo{person}{Zitao Liu}, \bibinfo{person}{Olga~C. Santos}, {and} \bibinfo{person}{Ig~Ibert Bittencourt}} (Eds.). \bibinfo{publisher}{Springer Nature Switzerland}, \bibinfo{address}{Cham}, \bibinfo{pages}{259--267}.
\newblock
\showISBNx{978-3-031-64299-9}


\bibitem[GitHub(2024)]%
        {github_copilot}
\bibfield{author}{\bibinfo{person}{GitHub}.} \bibinfo{year}{2024}\natexlab{}.
\newblock \bibinfo{title}{GitHub Copilot: Your AI Pair Programmer}.
\newblock
\newblock
\urldef\tempurl%
\url{https://github.com/features/copilot}
\showURL{%
\tempurl}
\newblock
\shownote{Accessed: 2024-09-01}.


\bibitem[Goddard et~al\mbox{.}(2012)]%
        {goddard2012automation}
\bibfield{author}{\bibinfo{person}{Kate Goddard}, \bibinfo{person}{Abdul Roudsari}, {and} \bibinfo{person}{Jeremy~C Wyatt}.} \bibinfo{year}{2012}\natexlab{}.
\newblock \showarticletitle{Automation bias: a systematic review of frequency, effect mediators, and mitigators}.
\newblock \bibinfo{journal}{\emph{Journal of the American Medical Informatics Association}} \bibinfo{volume}{19}, \bibinfo{number}{1} (\bibinfo{year}{2012}), \bibinfo{pages}{121--127}.
\newblock


\bibitem[Griffin(2019)]%
        {griffin2019bugs4learning}
\bibfield{author}{\bibinfo{person}{Jean~M. Griffin}.} \bibinfo{year}{2019}\natexlab{}.
\newblock \showarticletitle{Designing Intentional Bugs for Learning}. In \bibinfo{booktitle}{\emph{Proceedings of the 2019 Conference on United Kingdom \& Ireland Computing Education Research}} (Canterbury, United Kingdom) \emph{(\bibinfo{series}{UKICER '19})}. \bibinfo{publisher}{Association for Computing Machinery}, \bibinfo{address}{New York, NY, USA}, Article \bibinfo{articleno}{5}, \bibinfo{numpages}{7}~pages.
\newblock
\showISBNx{9781450372572}
\urldef\tempurl%
\url{https://doi.org/10.1145/3351287.3351289}
\showDOI{\tempurl}


\bibitem[Groe and Renkl(2007)]%
        {groe2007fixing_worked_examples}
\bibfield{author}{\bibinfo{person}{Cornelia~S Groe} {and} \bibinfo{person}{Alexander Renkl}.} \bibinfo{year}{2007}\natexlab{}.
\newblock \showarticletitle{Finding and fixing errors in worked examples: Can this foster learning outcomes?}
\newblock \bibinfo{journal}{\emph{Learning and Instruction}} \bibinfo{volume}{17}, \bibinfo{number}{6} (\bibinfo{year}{2007}), \bibinfo{pages}{612--634}.
\newblock
\showISSN{0959-4752}
\urldef\tempurl%
\url{https://doi.org/10.1016/j.learninstruc.2007.09.008}
\showDOI{\tempurl}


\bibitem[Gu et~al\mbox{.}(2024a)]%
        {gu2024analysts_ai_assistance}
\bibfield{author}{\bibinfo{person}{Ken Gu}, \bibinfo{person}{Madeleine Grunde-McLaughlin}, \bibinfo{person}{Andrew McNutt}, \bibinfo{person}{Jeffrey Heer}, {and} \bibinfo{person}{Tim Althoff}.} \bibinfo{year}{2024}\natexlab{a}.
\newblock \showarticletitle{How Do Data Analysts Respond to AI Assistance? A Wizard-of-Oz Study}. In \bibinfo{booktitle}{\emph{Proceedings of the 2024 CHI Conference on Human Factors in Computing Systems}} (Honolulu, HI, USA) \emph{(\bibinfo{series}{CHI '24})}. \bibinfo{publisher}{Association for Computing Machinery}, \bibinfo{address}{New York, NY, USA}, Article \bibinfo{articleno}{1015}, \bibinfo{numpages}{22}~pages.
\newblock
\showISBNx{9798400703300}
\urldef\tempurl%
\url{https://doi.org/10.1145/3613904.3641891}
\showDOI{\tempurl}


\bibitem[Gu et~al\mbox{.}(2024b)]%
        {gu2024verify_ai_assistance}
\bibfield{author}{\bibinfo{person}{Ken Gu}, \bibinfo{person}{Ruoxi Shang}, \bibinfo{person}{Tim Althoff}, \bibinfo{person}{Chenglong Wang}, {and} \bibinfo{person}{Steven~M. Drucker}.} \bibinfo{year}{2024}\natexlab{b}.
\newblock \showarticletitle{How Do Analysts Understand and Verify AI-Assisted Data Analyses?}. In \bibinfo{booktitle}{\emph{Proceedings of the 2024 CHI Conference on Human Factors in Computing Systems}} (Honolulu, HI, USA) \emph{(\bibinfo{series}{CHI '24})}. \bibinfo{publisher}{Association for Computing Machinery}, \bibinfo{address}{New York, NY, USA}, Article \bibinfo{articleno}{748}, \bibinfo{numpages}{22}~pages.
\newblock
\showISBNx{9798400703300}
\urldef\tempurl%
\url{https://doi.org/10.1145/3613904.3642497}
\showDOI{\tempurl}


\bibitem[Guo(2013a)]%
        {Guo2013SIGCSE}
\bibfield{author}{\bibinfo{person}{Philip~J. Guo}.} \bibinfo{year}{2013}\natexlab{a}.
\newblock \showarticletitle{Online Python Tutor: Embeddable Web-based Program Visualization for Cs Education}. In \bibinfo{booktitle}{\emph{Proceeding of the 44th ACM Technical Symposium on Computer Science Education}} (Denver, Colorado, USA) \emph{(\bibinfo{series}{SIGCSE '13})}. \bibinfo{publisher}{ACM}, \bibinfo{address}{New York, NY, USA}, \bibinfo{pages}{579--584}.
\newblock
\showISBNx{978-1-4503-1868-6}
\urldef\tempurl%
\url{https://doi.org/10.1145/2445196.2445368}
\showDOI{\tempurl}


\bibitem[Guo(2013b)]%
        {guo2013pythontutor}
\bibfield{author}{\bibinfo{person}{Philip~J. Guo}.} \bibinfo{year}{2013}\natexlab{b}.
\newblock \showarticletitle{Online python tutor: embeddable web-based program visualization for cs education}. In \bibinfo{booktitle}{\emph{Proceeding of the 44th ACM Technical Symposium on Computer Science Education}} (Denver, Colorado, USA) \emph{(\bibinfo{series}{SIGCSE '13})}. \bibinfo{publisher}{Association for Computing Machinery}, \bibinfo{address}{New York, NY, USA}, \bibinfo{pages}{579–584}.
\newblock
\showISBNx{9781450318686}
\urldef\tempurl%
\url{https://doi.org/10.1145/2445196.2445368}
\showDOI{\tempurl}


\bibitem[Guo(2023)]%
        {guo2023six_opportunities_chatgpt}
\bibfield{author}{\bibinfo{person}{Philip~J. Guo}.} \bibinfo{year}{2023}\natexlab{}.
\newblock \showarticletitle{Six Opportunities for Scientists and Engineers to Learn Programming Using AI Tools Such as ChatGPT}.
\newblock \bibinfo{journal}{\emph{Computing in Science \& Engineering}} \bibinfo{volume}{25}, \bibinfo{number}{3} (\bibinfo{year}{2023}), \bibinfo{pages}{73--78}.
\newblock
\urldef\tempurl%
\url{https://doi.org/10.1109/MCSE.2023.3308476}
\showDOI{\tempurl}


\bibitem[Hamouda et~al\mbox{.}(2017)]%
        {hamouda2017basic}
\bibfield{author}{\bibinfo{person}{Sally Hamouda}, \bibinfo{person}{Stephen~H Edwards}, \bibinfo{person}{Hicham~G Elmongui}, \bibinfo{person}{Jeremy~V Ernst}, {and} \bibinfo{person}{Clifford~A Shaffer}.} \bibinfo{year}{2017}\natexlab{}.
\newblock \showarticletitle{A basic recursion concept inventory}.
\newblock \bibinfo{journal}{\emph{Computer Science Education}} \bibinfo{volume}{27}, \bibinfo{number}{2} (\bibinfo{year}{2017}), \bibinfo{pages}{121--148}.
\newblock
\urldef\tempurl%
\url{https://doi.org/10.1080/08993408.2017.1414728}
\showDOI{\tempurl}
\showeprint{https://doi.org/10.1080/08993408.2017.1414728}


\bibitem[Harms et~al\mbox{.}(2013)]%
        {Harms2013IDC}
\bibfield{author}{\bibinfo{person}{Kyle~J. Harms}, \bibinfo{person}{Dennis Cosgrove}, \bibinfo{person}{Shannon Gray}, {and} \bibinfo{person}{Caitlin Kelleher}.} \bibinfo{year}{2013}\natexlab{}.
\newblock \showarticletitle{Automatically Generating Tutorials to Enable Middle School Children to Learn Programming Independently}. In \bibinfo{booktitle}{\emph{Proceedings of the 12th International Conference on Interaction Design and Children}} (New York, New York, USA) \emph{(\bibinfo{series}{IDC '13})}. \bibinfo{publisher}{ACM}, \bibinfo{address}{New York, NY, USA}, \bibinfo{pages}{11--19}.
\newblock
\showISBNx{978-1-4503-1918-8}
\urldef\tempurl%
\url{https://doi.org/10.1145/2485760.2485764}
\showDOI{\tempurl}


\bibitem[Hartmann et~al\mbox{.}(2010)]%
        {Hartmann2010CHI}
\bibfield{author}{\bibinfo{person}{Bj\"{o}rn Hartmann}, \bibinfo{person}{Daniel MacDougall}, \bibinfo{person}{Joel Brandt}, {and} \bibinfo{person}{Scott~R. Klemmer}.} \bibinfo{year}{2010}\natexlab{}.
\newblock \showarticletitle{What Would Other Programmers Do: Suggesting Solutions to Error Messages}. In \bibinfo{booktitle}{\emph{Proceedings of the SIGCHI Conference on Human Factors in Computing Systems}} (Atlanta, Georgia, USA) \emph{(\bibinfo{series}{CHI '10})}. \bibinfo{publisher}{ACM}, \bibinfo{address}{New York, NY, USA}, \bibinfo{pages}{1019--1028}.
\newblock
\showISBNx{978-1-60558-929-9}
\urldef\tempurl%
\url{https://doi.org/10.1145/1753326.1753478}
\showDOI{\tempurl}


\bibitem[Head et~al\mbox{.}(2015)]%
        {Head2015VLHCC}
\bibfield{author}{\bibinfo{person}{A. Head}, \bibinfo{person}{C. Appachu}, \bibinfo{person}{M.~A. Hearst}, {and} \bibinfo{person}{B. Hartmann}.} \bibinfo{year}{2015}\natexlab{}.
\newblock \showarticletitle{Tutorons: Generating context-relevant, on-demand explanations and demonstrations of online code}. In \bibinfo{booktitle}{\emph{2015 IEEE Symposium on Visual Languages and Human-Centric Computing (VL/HCC)}}. \bibinfo{pages}{3--12}.
\newblock
\urldef\tempurl%
\url{https://doi.org/10.1109/VLHCC.2015.7356972}
\showDOI{\tempurl}


\bibitem[Hoffswell et~al\mbox{.}(2018)]%
        {Hoffswell2018CHI}
\bibfield{author}{\bibinfo{person}{Jane Hoffswell}, \bibinfo{person}{Arvind Satyanarayan}, {and} \bibinfo{person}{Jeffrey Heer}.} \bibinfo{year}{2018}\natexlab{}.
\newblock \showarticletitle{Augmenting Code with In Situ Visualizations to Aid Program Understanding}. In \bibinfo{booktitle}{\emph{Proceedings of the 2018 CHI Conference on Human Factors in Computing Systems}} (Montreal QC, Canada) \emph{(\bibinfo{series}{CHI '18})}. \bibinfo{publisher}{ACM}, \bibinfo{address}{New York, NY, USA}, Article \bibinfo{articleno}{532}, \bibinfo{numpages}{12}~pages.
\newblock
\showISBNx{978-1-4503-5620-6}
\urldef\tempurl%
\url{https://doi.org/10.1145/3173574.3174106}
\showDOI{\tempurl}


\bibitem[Hofman et~al\mbox{.}(2023)]%
        {hofman2023sports_ai}
\bibfield{author}{\bibinfo{person}{Jake Hofman}, \bibinfo{person}{Daniel~G. Goldstein}, {and} \bibinfo{person}{David Rothschild}.} \bibinfo{year}{2023}\natexlab{}.
\newblock \showarticletitle{A Sports Analogy for Understanding Different Ways to Use AI}.
\newblock \bibinfo{journal}{\emph{Harvard Business Review}} (\bibinfo{date}{December} \bibinfo{year}{2023}).
\newblock
\urldef\tempurl%
\url{https://www.microsoft.com/en-us/research/publication/a-sports-analogy-for-understanding-different-ways-to-use-ai/}
\showURL{%
\tempurl}


\bibitem[Hou et~al\mbox{.}(2024a)]%
        {hou2024ai_solve_parsons}
\bibfield{author}{\bibinfo{person}{Irene Hou}, \bibinfo{person}{Owen Man}, \bibinfo{person}{Sophia Mettille}, \bibinfo{person}{Sebastian Gutierrez}, \bibinfo{person}{Kenneth Angelikas}, {and} \bibinfo{person}{Stephen MacNeil}.} \bibinfo{year}{2024}\natexlab{a}.
\newblock \showarticletitle{More Robots are Coming: Large Multimodal Models (ChatGPT) can Solve Visually Diverse Images of Parsons Problems}. In \bibinfo{booktitle}{\emph{Proceedings of the 26th Australasian Computing Education Conference}} (Sydney, NSW, Australia) \emph{(\bibinfo{series}{ACE '24})}. \bibinfo{publisher}{Association for Computing Machinery}, \bibinfo{address}{New York, NY, USA}, \bibinfo{pages}{29–38}.
\newblock
\showISBNx{9798400716195}
\urldef\tempurl%
\url{https://doi.org/10.1145/3636243.3636247}
\showDOI{\tempurl}


\bibitem[Hou et~al\mbox{.}(2024b)]%
        {hou2024effects}
\bibfield{author}{\bibinfo{person}{Irene Hou}, \bibinfo{person}{Sophia Mettille}, \bibinfo{person}{Owen Man}, \bibinfo{person}{Zhuo Li}, \bibinfo{person}{Cynthia Zastudil}, {and} \bibinfo{person}{Stephen MacNeil}.} \bibinfo{year}{2024}\natexlab{b}.
\newblock \showarticletitle{The Effects of Generative AI on Computing Students' Help-Seeking Preferences}. In \bibinfo{booktitle}{\emph{Proceedings of the 26th Australasian Computing Education Conference}} (Sydney, NSW, Australia) \emph{(\bibinfo{series}{ACE '24})}. \bibinfo{publisher}{Association for Computing Machinery}, \bibinfo{address}{New York, NY, USA}, \bibinfo{pages}{39–48}.
\newblock
\showISBNx{9798400716195}
\urldef\tempurl%
\url{https://doi.org/10.1145/3636243.3636248}
\showDOI{\tempurl}


\bibitem[Hou et~al\mbox{.}(2024c)]%
        {hou2024codetailor}
\bibfield{author}{\bibinfo{person}{Xinying Hou}, \bibinfo{person}{Zihan Wu}, \bibinfo{person}{Xu Wang}, {and} \bibinfo{person}{Barbara~J. Ericson}.} \bibinfo{year}{2024}\natexlab{c}.
\newblock \showarticletitle{CodeTailor: LLM-Powered Personalized Parsons Puzzles for Engaging Support While Learning Programming}. In \bibinfo{booktitle}{\emph{Proceedings of the Eleventh ACM Conference on Learning @ Scale}} (Atlanta, GA, USA) \emph{(\bibinfo{series}{L@S '24})}. \bibinfo{publisher}{Association for Computing Machinery}, \bibinfo{address}{New York, NY, USA}, \bibinfo{pages}{51–62}.
\newblock
\showISBNx{9798400706332}
\urldef\tempurl%
\url{https://doi.org/10.1145/3657604.3662032}
\showDOI{\tempurl}


\bibitem[Hristova et~al\mbox{.}(2003)]%
        {Hristova2003SIGCSE}
\bibfield{author}{\bibinfo{person}{Maria Hristova}, \bibinfo{person}{Ananya Misra}, \bibinfo{person}{Megan Rutter}, {and} \bibinfo{person}{Rebecca Mercuri}.} \bibinfo{year}{2003}\natexlab{}.
\newblock \showarticletitle{Identifying and correcting Java programming errors for introductory computer science students}. In \bibinfo{booktitle}{\emph{Proceedings of the 34th SIGCSE Technical Symposium on Computer Science Education}} (Reno, Navada, USA) \emph{(\bibinfo{series}{SIGCSE '03})}. \bibinfo{publisher}{Association for Computing Machinery}, \bibinfo{address}{New York, NY, USA}, \bibinfo{pages}{153–156}.
\newblock
\showISBNx{158113648X}
\urldef\tempurl%
\url{https://doi.org/10.1145/611892.611956}
\showDOI{\tempurl}


\bibitem[Ichinco et~al\mbox{.}(2016)]%
        {Ichinco2016VLHCC}
\bibfield{author}{\bibinfo{person}{M. Ichinco}, \bibinfo{person}{W. Hnin}, {and} \bibinfo{person}{C. Kelleher}.} \bibinfo{year}{2016}\natexlab{}.
\newblock \showarticletitle{Suggesting examples to novice programmers in an open-ended context with the example guru}. In \bibinfo{booktitle}{\emph{2016 IEEE Symposium on Visual Languages and Human-Centric Computing (VL/HCC)}}. \bibinfo{pages}{230--231}.
\newblock
\showISSN{1943-6106}
\urldef\tempurl%
\url{https://doi.org/10.1109/VLHCC.2016.7739691}
\showDOI{\tempurl}


\bibitem[Ichinco et~al\mbox{.}(2017)]%
        {Ichinco2017CHI}
\bibfield{author}{\bibinfo{person}{Michelle Ichinco}, \bibinfo{person}{Wint~Yee Hnin}, {and} \bibinfo{person}{Caitlin~L. Kelleher}.} \bibinfo{year}{2017}\natexlab{}.
\newblock \showarticletitle{Suggesting API Usage to Novice Programmers with the Example Guru}. In \bibinfo{booktitle}{\emph{Proceedings of the 2017 CHI Conference on Human Factors in Computing Systems}} (Denver, Colorado, USA) \emph{(\bibinfo{series}{CHI '17})}. \bibinfo{publisher}{ACM}, \bibinfo{address}{New York, NY, USA}, \bibinfo{pages}{1105--1117}.
\newblock
\showISBNx{978-1-4503-4655-9}
\urldef\tempurl%
\url{https://doi.org/10.1145/3025453.3025827}
\showDOI{\tempurl}


\bibitem[Ichinco and Kelleher(2015)]%
        {ichinco2015exploring}
\bibfield{author}{\bibinfo{person}{Michelle Ichinco} {and} \bibinfo{person}{Caitlin Kelleher}.} \bibinfo{year}{2015}\natexlab{}.
\newblock \showarticletitle{Exploring novice programmer example use}. In \bibinfo{booktitle}{\emph{2015 IEEE Symposium on Visual Languages and Human-Centric Computing (VL/HCC)}}. \bibinfo{pages}{63--71}.
\newblock
\urldef\tempurl%
\url{https://doi.org/10.1109/VLHCC.2015.7357199}
\showDOI{\tempurl}


\bibitem[Ichinco and Kelleher(2018)]%
        {Ichinco2018IDC}
\bibfield{author}{\bibinfo{person}{Michelle Ichinco} {and} \bibinfo{person}{Caitlin Kelleher}.} \bibinfo{year}{2018}\natexlab{}.
\newblock \showarticletitle{Semi-automatic Suggestion Generation for Young Novice Programmers in an Open-ended Context}. In \bibinfo{booktitle}{\emph{Proceedings of the 17th ACM Conference on Interaction Design and Children}} (Trondheim, Norway) \emph{(\bibinfo{series}{IDC '18})}. \bibinfo{publisher}{ACM}, \bibinfo{address}{New York, NY, USA}, \bibinfo{pages}{405--412}.
\newblock
\showISBNx{978-1-4503-5152-2}
\urldef\tempurl%
\url{https://doi.org/10.1145/3202185.3202762}
\showDOI{\tempurl}


\bibitem[Jackson et~al\mbox{.}(2005)]%
        {Jackson2005FIE}
\bibfield{author}{\bibinfo{person}{J. Jackson}, \bibinfo{person}{M. Cobb}, {and} \bibinfo{person}{C. Carver}.} \bibinfo{year}{2005}\natexlab{}.
\newblock \showarticletitle{Identifying Top Java Errors for Novice Programmers}. In \bibinfo{booktitle}{\emph{Proceedings Frontiers in Education 35th Annual Conference}}. \bibinfo{pages}{T4C--T4C}.
\newblock
\urldef\tempurl%
\url{https://doi.org/10.1109/FIE.2005.1611967}
\showDOI{\tempurl}


\bibitem[Kaczmarczyk et~al\mbox{.}(2010)]%
        {kaczmarczyk2010identifying}
\bibfield{author}{\bibinfo{person}{Lisa~C. Kaczmarczyk}, \bibinfo{person}{Elizabeth~R. Petrick}, \bibinfo{person}{J.~Philip East}, {and} \bibinfo{person}{Geoffrey~L. Herman}.} \bibinfo{year}{2010}\natexlab{}.
\newblock \showarticletitle{Identifying student misconceptions of programming}. In \bibinfo{booktitle}{\emph{Proceedings of the 41st ACM Technical Symposium on Computer Science Education}} (Milwaukee, Wisconsin, USA) \emph{(\bibinfo{series}{SIGCSE '10})}. \bibinfo{publisher}{Association for Computing Machinery}, \bibinfo{address}{New York, NY, USA}, \bibinfo{pages}{107–111}.
\newblock
\showISBNx{9781450300063}
\urldef\tempurl%
\url{https://doi.org/10.1145/1734263.1734299}
\showDOI{\tempurl}


\bibitem[Kang and Guo(2017)]%
        {Kang2017UIST}
\bibfield{author}{\bibinfo{person}{Hyeonsu Kang} {and} \bibinfo{person}{Philip~J. Guo}.} \bibinfo{year}{2017}\natexlab{}.
\newblock \showarticletitle{Omnicode: A Novice-Oriented Live Programming Environment with Always-On Run-Time Value Visualizations}. In \bibinfo{booktitle}{\emph{Proceedings of the 30th Annual ACM Symposium on User Interface Software and Technology}} (Qu\&\#233;bec City, QC, Canada) \emph{(\bibinfo{series}{UIST '17})}. \bibinfo{publisher}{ACM}, \bibinfo{address}{New York, NY, USA}, \bibinfo{pages}{737--745}.
\newblock
\showISBNx{978-1-4503-4981-9}
\urldef\tempurl%
\url{https://doi.org/10.1145/3126594.3126632}
\showDOI{\tempurl}


\bibitem[Kasneci et~al\mbox{.}(2023)]%
        {Kasneci2023ChatGPT}
\bibfield{author}{\bibinfo{person}{Enkelejda Kasneci}, \bibinfo{person}{Kathrin Sessler}, \bibinfo{person}{Stefan Küchemann}, \bibinfo{person}{Maria Bannert}, \bibinfo{person}{Daryna Dementieva}, \bibinfo{person}{Frank Fischer}, \bibinfo{person}{Urs Gasser}, \bibinfo{person}{Georg Groh}, \bibinfo{person}{Stephan Günnemann}, \bibinfo{person}{Eyke Hüllermeier}, \bibinfo{person}{Stephan Krusche}, \bibinfo{person}{Gitta Kutyniok}, \bibinfo{person}{Tilman Michaeli}, \bibinfo{person}{Claudia Nerdel}, \bibinfo{person}{Jürgen Pfeffer}, \bibinfo{person}{Oleksandra Poquet}, \bibinfo{person}{Michael Sailer}, \bibinfo{person}{Albrecht Schmidt}, \bibinfo{person}{Tina Seidel}, \bibinfo{person}{Matthias Stadler}, \bibinfo{person}{Jochen Weller}, \bibinfo{person}{Jochen Kuhn}, {and} \bibinfo{person}{Gjergji Kasneci}.} \bibinfo{year}{2023}\natexlab{}.
\newblock \showarticletitle{ChatGPT for good? On opportunities and challenges of large language models for education}.
\newblock \bibinfo{journal}{\emph{Learning and Individual Differences}}  \bibinfo{volume}{103} (\bibinfo{year}{2023}), \bibinfo{pages}{102274}.
\newblock
\showISSN{1041-6080}
\urldef\tempurl%
\url{https://doi.org/10.1016/j.lindif.2023.102274}
\showDOI{\tempurl}


\bibitem[Kazemitabaar et~al\mbox{.}(2023)]%
        {kazemitabaar_chi23_effect_ai}
\bibfield{author}{\bibinfo{person}{Majeed Kazemitabaar}, \bibinfo{person}{Justin Chow}, \bibinfo{person}{Carl Ka~To Ma}, \bibinfo{person}{Barbara~J. Ericson}, \bibinfo{person}{David Weintrop}, {and} \bibinfo{person}{Tovi Grossman}.} \bibinfo{year}{2023}\natexlab{}.
\newblock \showarticletitle{Studying the effect of AI Code Generators on Supporting Novice Learners in Introductory Programming}. In \bibinfo{booktitle}{\emph{Proceedings of the 2023 CHI Conference on Human Factors in Computing Systems}} (Hamburg, Germany) \emph{(\bibinfo{series}{CHI '23})}. \bibinfo{publisher}{Association for Computing Machinery}, \bibinfo{address}{New York, NY, USA}, \bibinfo{numpages}{23}~pages.
\newblock
\showISBNx{9781450394215}
\urldef\tempurl%
\url{https://doi.org/10.1145/3544548.3580919}
\showDOI{\tempurl}


\bibitem[Kazemitabaar et~al\mbox{.}(2024a)]%
        {kazemitabaar_koli23_using_ai}
\bibfield{author}{\bibinfo{person}{Majeed Kazemitabaar}, \bibinfo{person}{Xinying Hou}, \bibinfo{person}{Austin Henley}, \bibinfo{person}{Barbara~Jane Ericson}, \bibinfo{person}{David Weintrop}, {and} \bibinfo{person}{Tovi Grossman}.} \bibinfo{year}{2024}\natexlab{a}.
\newblock \showarticletitle{How Novices Use LLM-based Code Generators to Solve CS1 Coding Tasks in a Self-Paced Learning Environment}. In \bibinfo{booktitle}{\emph{Proceedings of the 23rd Koli Calling International Conference on Computing Education Research}} (Koli, Finland) \emph{(\bibinfo{series}{Koli Calling '23})}. \bibinfo{publisher}{Association for Computing Machinery}, \bibinfo{address}{New York, NY, USA}, \bibinfo{numpages}{12}~pages.
\newblock
\showISBNx{9798400716539}
\urldef\tempurl%
\url{https://doi.org/10.1145/3631802.3631806}
\showDOI{\tempurl}


\bibitem[Kazemitabaar et~al\mbox{.}(2024b)]%
        {kazemitabaar_uist24_steering_verification}
\bibfield{author}{\bibinfo{person}{Majeed Kazemitabaar}, \bibinfo{person}{Jack Williams}, \bibinfo{person}{Ian Drosos}, \bibinfo{person}{Tovi Grossman}, \bibinfo{person}{Austin Henley}, \bibinfo{person}{Carina Negreanu}, {and} \bibinfo{person}{Advait Sarkar}.} \bibinfo{year}{2024}\natexlab{b}.
\newblock \showarticletitle{Improving Steering and Verification in AI-Assisted Data Analysis with Interactive Task Decomposition}. In \bibinfo{booktitle}{\emph{Proceedings of the 37th Annual ACM Symposium on User Interface Software and Technology}} (Pittsburgh, PA, USA) \emph{(\bibinfo{series}{UIST '24})}. \bibinfo{publisher}{Association for Computing Machinery}, \bibinfo{address}{New York, NY, USA}, \bibinfo{numpages}{19}~pages.
\newblock
\urldef\tempurl%
\url{https://doi.org/10.1145/3654777.3676345}
\showDOI{\tempurl}


\bibitem[Kazemitabaar et~al\mbox{.}(2024c)]%
        {kazemitabaar_chi24_codeaid}
\bibfield{author}{\bibinfo{person}{Majeed Kazemitabaar}, \bibinfo{person}{Runlong Ye}, \bibinfo{person}{Xiaoning Wang}, \bibinfo{person}{Austin~Zachary Henley}, \bibinfo{person}{Paul Denny}, \bibinfo{person}{Michelle Craig}, {and} \bibinfo{person}{Tovi Grossman}.} \bibinfo{year}{2024}\natexlab{c}.
\newblock \showarticletitle{CodeAid: Evaluating a Classroom Deployment of an LLM-based Programming Assistant that Balances Student and Educator Needs}. In \bibinfo{booktitle}{\emph{Proceedings of the CHI Conference on Human Factors in Computing Systems}} (Honolulu, HI, USA) \emph{(\bibinfo{series}{CHI '24})}. \bibinfo{publisher}{Association for Computing Machinery}, \bibinfo{address}{New York, NY, USA}, Article \bibinfo{articleno}{650}, \bibinfo{numpages}{20}~pages.
\newblock
\showISBNx{9798400703300}
\urldef\tempurl%
\url{https://doi.org/10.1145/3613904.3642773}
\showDOI{\tempurl}


\bibitem[Kennedy and Kraemer(2019)]%
        {Kennedy2019ITiCSE}
\bibfield{author}{\bibinfo{person}{Cazembe Kennedy} {and} \bibinfo{person}{Eileen~T. Kraemer}.} \bibinfo{year}{2019}\natexlab{}.
\newblock \showarticletitle{Qualitative Observations of Student Reasoning: Coding in the Wild}. In \bibinfo{booktitle}{\emph{Proceedings of the 2019 ACM Conference on Innovation and Technology in Computer Science Education}} (Aberdeen, Scotland Uk) \emph{(\bibinfo{series}{ITiCSE '19})}. \bibinfo{publisher}{Association for Computing Machinery}, \bibinfo{address}{New York, NY, USA}, \bibinfo{pages}{224–230}.
\newblock
\showISBNx{9781450368957}
\urldef\tempurl%
\url{https://doi.org/10.1145/3304221.3319751}
\showDOI{\tempurl}


\bibitem[Ko et~al\mbox{.}(2004)]%
        {ko2004learning_barriers_eup}
\bibfield{author}{\bibinfo{person}{Amy~J. Ko}, \bibinfo{person}{Brad~A. Myers}, {and} \bibinfo{person}{Htet~Htet Aung}.} \bibinfo{year}{2004}\natexlab{}.
\newblock \showarticletitle{Six Learning Barriers in End-User Programming Systems}. In \bibinfo{booktitle}{\emph{2004 IEEE Symposium on Visual Languages - Human Centric Computing}}. \bibinfo{pages}{199--206}.
\newblock
\urldef\tempurl%
\url{https://doi.org/10.1109/VLHCC.2004.47}
\showDOI{\tempurl}


\bibitem[Kolikant(2005)]%
        {Kolikant2005ICER}
\bibfield{author}{\bibinfo{person}{Yifat Ben-David Kolikant}.} \bibinfo{year}{2005}\natexlab{}.
\newblock \showarticletitle{Students' alternative standards for correctness}. In \bibinfo{booktitle}{\emph{Proceedings of the First International Workshop on Computing Education Research}} (Seattle, WA, USA) \emph{(\bibinfo{series}{ICER '05})}. \bibinfo{publisher}{Association for Computing Machinery}, \bibinfo{address}{New York, NY, USA}, \bibinfo{pages}{37–43}.
\newblock
\showISBNx{1595930434}
\urldef\tempurl%
\url{https://doi.org/10.1145/1089786.1089790}
\showDOI{\tempurl}


\bibitem[Kosch et~al\mbox{.}(2023)]%
        {kosch2023nasa_tlx}
\bibfield{author}{\bibinfo{person}{Thomas Kosch}, \bibinfo{person}{Jakob Karolus}, \bibinfo{person}{Johannes Zagermann}, \bibinfo{person}{Harald Reiterer}, \bibinfo{person}{Albrecht Schmidt}, {and} \bibinfo{person}{Pawe\l{}~W. Wo\'{z}niak}.} \bibinfo{year}{2023}\natexlab{}.
\newblock \showarticletitle{A Survey on Measuring Cognitive Workload in Human-Computer Interaction}.
\newblock \bibinfo{journal}{\emph{ACM Comput. Surv.}} \bibinfo{volume}{55}, \bibinfo{number}{13s}, Article \bibinfo{articleno}{283} (\bibinfo{date}{July} \bibinfo{year}{2023}), \bibinfo{numpages}{39}~pages.
\newblock
\showISSN{0360-0300}
\urldef\tempurl%
\url{https://doi.org/10.1145/3582272}
\showDOI{\tempurl}


\bibitem[Kumar(2013)]%
        {kumar2013tracing_writing}
\bibfield{author}{\bibinfo{person}{Amruth~N. Kumar}.} \bibinfo{year}{2013}\natexlab{}.
\newblock \showarticletitle{A study of the influence of code-tracing problems on code-writing skills}. In \bibinfo{booktitle}{\emph{Proceedings of the 18th ACM Conference on Innovation and Technology in Computer Science Education}} (Canterbury, England, UK) \emph{(\bibinfo{series}{ITiCSE '13})}. \bibinfo{publisher}{Association for Computing Machinery}, \bibinfo{address}{New York, NY, USA}, \bibinfo{pages}{183–188}.
\newblock
\showISBNx{9781450320788}
\urldef\tempurl%
\url{https://doi.org/10.1145/2462476.2462507}
\showDOI{\tempurl}


\bibitem[Lahtinen et~al\mbox{.}(2005)]%
        {lahtinen2005difficulties}
\bibfield{author}{\bibinfo{person}{Essi Lahtinen}, \bibinfo{person}{Kirsti Ala-Mutka}, {and} \bibinfo{person}{Hannu-Matti J\"{a}rvinen}.} \bibinfo{year}{2005}\natexlab{}.
\newblock \showarticletitle{A study of the difficulties of novice programmers}.
\newblock \bibinfo{journal}{\emph{SIGCSE Bull.}} \bibinfo{volume}{37}, \bibinfo{number}{3} (\bibinfo{date}{jun} \bibinfo{year}{2005}), \bibinfo{pages}{14–18}.
\newblock
\showISSN{0097-8418}
\urldef\tempurl%
\url{https://doi.org/10.1145/1151954.1067453}
\showDOI{\tempurl}


\bibitem[Lau and Guo(2023)]%
        {lau2023ban}
\bibfield{author}{\bibinfo{person}{Sam Lau} {and} \bibinfo{person}{Philip Guo}.} \bibinfo{year}{2023}\natexlab{}.
\newblock \showarticletitle{From "Ban It Till We Understand It" to "Resistance is Futile": How University Programming Instructors Plan to Adapt as More Students Use AI Code Generation and Explanation Tools such as ChatGPT and GitHub Copilot}. In \bibinfo{booktitle}{\emph{Proceedings of the 2023 ACM Conference on International Computing Education Research - Volume 1}} (Chicago, IL, USA) \emph{(\bibinfo{series}{ICER '23})}. \bibinfo{publisher}{Association for Computing Machinery}, \bibinfo{address}{New York, NY, USA}, \bibinfo{pages}{106–121}.
\newblock
\showISBNx{9781450399760}
\urldef\tempurl%
\url{https://doi.org/10.1145/3568813.3600138}
\showDOI{\tempurl}


\bibitem[Lee et~al\mbox{.}(2014)]%
        {lee2014debug_learn_code}
\bibfield{author}{\bibinfo{person}{Michael~J. Lee}, \bibinfo{person}{Faezeh Bahmani}, \bibinfo{person}{Irwin Kwan}, \bibinfo{person}{Jilian LaFerte}, \bibinfo{person}{Polina Charters}, \bibinfo{person}{Amber Horvath}, \bibinfo{person}{Fanny Luor}, \bibinfo{person}{Jill Cao}, \bibinfo{person}{Catherine Law}, \bibinfo{person}{Michael Beswetherick}, \bibinfo{person}{Sheridan Long}, \bibinfo{person}{Margaret Burnett}, {and} \bibinfo{person}{Amy~J. Ko}.} \bibinfo{year}{2014}\natexlab{}.
\newblock \showarticletitle{Principles of a debugging-first puzzle game for computing education}. In \bibinfo{booktitle}{\emph{2014 IEEE Symposium on Visual Languages and Human-Centric Computing (VL/HCC)}}. \bibinfo{pages}{57--64}.
\newblock
\urldef\tempurl%
\url{https://doi.org/10.1109/VLHCC.2014.6883023}
\showDOI{\tempurl}


\bibitem[Lieber et~al\mbox{.}(2014)]%
        {Lieber2014CHI}
\bibfield{author}{\bibinfo{person}{Tom Lieber}, \bibinfo{person}{Joel~R. Brandt}, {and} \bibinfo{person}{Rob~C. Miller}.} \bibinfo{year}{2014}\natexlab{}.
\newblock \showarticletitle{Addressing Misconceptions About Code with Always-on Programming Visualizations}. In \bibinfo{booktitle}{\emph{Proceedings of the SIGCHI Conference on Human Factors in Computing Systems}} (Toronto, Ontario, Canada) \emph{(\bibinfo{series}{CHI '14})}. \bibinfo{publisher}{ACM}, \bibinfo{address}{New York, NY, USA}, \bibinfo{pages}{2481--2490}.
\newblock
\showISBNx{978-1-4503-2473-1}
\urldef\tempurl%
\url{https://doi.org/10.1145/2556288.2557409}
\showDOI{\tempurl}


\bibitem[Liffiton et~al\mbox{.}(2024)]%
        {Liffiton2024codehelp}
\bibfield{author}{\bibinfo{person}{Mark Liffiton}, \bibinfo{person}{Brad~E Sheese}, \bibinfo{person}{Jaromir Savelka}, {and} \bibinfo{person}{Paul Denny}.} \bibinfo{year}{2024}\natexlab{}.
\newblock \showarticletitle{CodeHelp: Using Large Language Models with Guardrails for Scalable Support in Programming Classes}. In \bibinfo{booktitle}{\emph{Proceedings of the 23rd Koli Calling International Conference on Computing Education Research}} (Koli, Finland) \emph{(\bibinfo{series}{Koli Calling '23})}. \bibinfo{publisher}{Association for Computing Machinery}, \bibinfo{address}{New York, NY, USA}, Article \bibinfo{articleno}{8}, \bibinfo{numpages}{11}~pages.
\newblock
\showISBNx{9798400716539}
\urldef\tempurl%
\url{https://doi.org/10.1145/3631802.3631830}
\showDOI{\tempurl}


\bibitem[Linn and Clancy(1992)]%
        {linn1992case}
\bibfield{author}{\bibinfo{person}{Marcia~C Linn} {and} \bibinfo{person}{Michael~J Clancy}.} \bibinfo{year}{1992}\natexlab{}.
\newblock \showarticletitle{The case for case studies of programming problems}.
\newblock \bibinfo{journal}{\emph{Commun. ACM}} \bibinfo{volume}{35}, \bibinfo{number}{3} (\bibinfo{year}{1992}), \bibinfo{pages}{121--132}.
\newblock


\bibitem[Lister et~al\mbox{.}(2006)]%
        {Lister2006ITICSE}
\bibfield{author}{\bibinfo{person}{Raymond Lister}, \bibinfo{person}{Beth Simon}, \bibinfo{person}{Errol Thompson}, \bibinfo{person}{Jacqueline~L. Whalley}, {and} \bibinfo{person}{Christine Prasad}.} \bibinfo{year}{2006}\natexlab{}.
\newblock \showarticletitle{Not seeing the forest for the trees: novice programmers and the SOLO taxonomy}. In \bibinfo{booktitle}{\emph{Proceedings of the 11th Annual SIGCSE Conference on Innovation and Technology in Computer Science Education}} (Bologna, Italy) \emph{(\bibinfo{series}{ITICSE '06})}. \bibinfo{publisher}{Association for Computing Machinery}, \bibinfo{address}{New York, NY, USA}, \bibinfo{pages}{118–122}.
\newblock
\showISBNx{1595930558}
\urldef\tempurl%
\url{https://doi.org/10.1145/1140124.1140157}
\showDOI{\tempurl}


\bibitem[Lopez et~al\mbox{.}(2008)]%
        {lopez2008tracing_writing}
\bibfield{author}{\bibinfo{person}{Mike Lopez}, \bibinfo{person}{Jacqueline Whalley}, \bibinfo{person}{Phil Robbins}, {and} \bibinfo{person}{Raymond Lister}.} \bibinfo{year}{2008}\natexlab{}.
\newblock \showarticletitle{Relationships between reading, tracing and writing skills in introductory programming}. In \bibinfo{booktitle}{\emph{Proceedings of the Fourth International Workshop on Computing Education Research}} (Sydney, Australia) \emph{(\bibinfo{series}{ICER '08})}. \bibinfo{publisher}{Association for Computing Machinery}, \bibinfo{address}{New York, NY, USA}, \bibinfo{pages}{101–112}.
\newblock
\showISBNx{9781605582160}
\urldef\tempurl%
\url{https://doi.org/10.1145/1404520.1404531}
\showDOI{\tempurl}


\bibitem[Mayer(1981)]%
        {mayer1981psychology}
\bibfield{author}{\bibinfo{person}{Richard~E. Mayer}.} \bibinfo{year}{1981}\natexlab{}.
\newblock \showarticletitle{The Psychology of How Novices Learn Computer Programming}.
\newblock \bibinfo{journal}{\emph{ACM Comput. Surv.}} \bibinfo{volume}{13}, \bibinfo{number}{1} (\bibinfo{date}{mar} \bibinfo{year}{1981}), \bibinfo{pages}{121–141}.
\newblock
\showISSN{0360-0300}
\urldef\tempurl%
\url{https://doi.org/10.1145/356835.356841}
\showDOI{\tempurl}


\bibitem[McCauley et~al\mbox{.}(2008)]%
        {McCauley2008CSE}
\bibfield{author}{\bibinfo{person}{Renee McCauley}, \bibinfo{person}{Sue Fitzgerald}, \bibinfo{person}{Gary Lewandowski}, \bibinfo{person}{Laurie Murphy}, \bibinfo{person}{Beth Simon}, \bibinfo{person}{Lynda Thomas}, {and} \bibinfo{person}{Carol Zander}.} \bibinfo{year}{2008}\natexlab{}.
\newblock \showarticletitle{Debugging: a review of the literature from an educational perspective}.
\newblock \bibinfo{journal}{\emph{Computer Science Education}} \bibinfo{volume}{18}, \bibinfo{number}{2} (\bibinfo{year}{2008}), \bibinfo{pages}{67--92}.
\newblock
\urldef\tempurl%
\url{https://doi.org/10.1080/08993400802114581}
\showDOI{\tempurl}
\showeprint{https://doi.org/10.1080/08993400802114581}


\bibitem[Miedema et~al\mbox{.}(2022)]%
        {miedema2022identifying}
\bibfield{author}{\bibinfo{person}{Daphne Miedema}, \bibinfo{person}{Efthimia Aivaloglou}, {and} \bibinfo{person}{George Fletcher}.} \bibinfo{year}{2022}\natexlab{}.
\newblock \showarticletitle{Identifying SQL misconceptions of novices: findings from a think-aloud study}.
\newblock \bibinfo{journal}{\emph{ACM Inroads}} \bibinfo{volume}{13}, \bibinfo{number}{1} (\bibinfo{date}{feb} \bibinfo{year}{2022}), \bibinfo{pages}{52–65}.
\newblock
\showISSN{2153-2184}
\urldef\tempurl%
\url{https://doi.org/10.1145/3514214}
\showDOI{\tempurl}


\bibitem[Morrison et~al\mbox{.}(2015)]%
        {morrison2015workedexamples}
\bibfield{author}{\bibinfo{person}{Briana~B. Morrison}, \bibinfo{person}{Lauren~E. Margulieux}, {and} \bibinfo{person}{Mark Guzdial}.} \bibinfo{year}{2015}\natexlab{}.
\newblock \showarticletitle{Subgoals, Context, and Worked Examples in Learning Computing Problem Solving}. In \bibinfo{booktitle}{\emph{Proceedings of the Eleventh Annual International Conference on International Computing Education Research}} (Omaha, Nebraska, USA) \emph{(\bibinfo{series}{ICER '15})}. \bibinfo{publisher}{Association for Computing Machinery}, \bibinfo{address}{New York, NY, USA}, \bibinfo{pages}{21–29}.
\newblock
\showISBNx{9781450336307}
\urldef\tempurl%
\url{https://doi.org/10.1145/2787622.2787733}
\showDOI{\tempurl}


\bibitem[Mujumdar et~al\mbox{.}(2011)]%
        {Mujumdar2011CHI}
\bibfield{author}{\bibinfo{person}{Dhawal Mujumdar}, \bibinfo{person}{Manuel Kallenbach}, \bibinfo{person}{Brandon Liu}, {and} \bibinfo{person}{Bj\"{o}rn Hartmann}.} \bibinfo{year}{2011}\natexlab{}.
\newblock \showarticletitle{Crowdsourcing Suggestions to Programming Problems for Dynamic Web Development Languages}. In \bibinfo{booktitle}{\emph{CHI '11 Extended Abstracts on Human Factors in Computing Systems}} (Vancouver, BC, Canada) \emph{(\bibinfo{series}{CHI EA '11})}. \bibinfo{publisher}{ACM}, \bibinfo{address}{New York, NY, USA}, \bibinfo{pages}{1525--1530}.
\newblock
\showISBNx{978-1-4503-0268-5}
\urldef\tempurl%
\url{https://doi.org/10.1145/1979742.1979802}
\showDOI{\tempurl}


\bibitem[Murphy et~al\mbox{.}(2012)]%
        {murphy2012explain_writing_code}
\bibfield{author}{\bibinfo{person}{Laurie Murphy}, \bibinfo{person}{Sue Fitzgerald}, \bibinfo{person}{Raymond Lister}, {and} \bibinfo{person}{Ren\'{e}e McCauley}.} \bibinfo{year}{2012}\natexlab{}.
\newblock \showarticletitle{Ability to 'explain in plain english' linked to proficiency in computer-based programming}. In \bibinfo{booktitle}{\emph{Proceedings of the Ninth Annual International Conference on International Computing Education Research}} (Auckland, New Zealand) \emph{(\bibinfo{series}{ICER '12})}. \bibinfo{publisher}{Association for Computing Machinery}, \bibinfo{address}{New York, NY, USA}, \bibinfo{pages}{111–118}.
\newblock
\showISBNx{9781450316040}
\urldef\tempurl%
\url{https://doi.org/10.1145/2361276.2361299}
\showDOI{\tempurl}


\bibitem[Nienaltowski et~al\mbox{.}(2008)]%
        {Nienaltowski2008SIGCSE}
\bibfield{author}{\bibinfo{person}{Marie-H{\'e}l\`{e}ne Nienaltowski}, \bibinfo{person}{Michela Pedroni}, {and} \bibinfo{person}{Bertrand Meyer}.} \bibinfo{year}{2008}\natexlab{}.
\newblock \showarticletitle{Compiler Error Messages: What Can Help Novices?}. In \bibinfo{booktitle}{\emph{Proceedings of the 39th SIGCSE Technical Symposium on Computer Science Education}} (Portland, OR, USA) \emph{(\bibinfo{series}{SIGCSE '08})}. \bibinfo{publisher}{ACM}, \bibinfo{address}{New York, NY, USA}, \bibinfo{pages}{168--172}.
\newblock
\showISBNx{978-1-59593-799-5}
\urldef\tempurl%
\url{https://doi.org/10.1145/1352135.1352192}
\showDOI{\tempurl}


\bibitem[Oney and Brandt(2012)]%
        {Oney2012CHI}
\bibfield{author}{\bibinfo{person}{Stephen Oney} {and} \bibinfo{person}{Joel Brandt}.} \bibinfo{year}{2012}\natexlab{}.
\newblock \showarticletitle{Codelets: Linking Interactive Documentation and Example Code in the Editor}. In \bibinfo{booktitle}{\emph{Proceedings of the SIGCHI Conference on Human Factors in Computing Systems}} (Austin, Texas, USA) \emph{(\bibinfo{series}{CHI '12})}. \bibinfo{publisher}{ACM}, \bibinfo{address}{New York, NY, USA}, \bibinfo{pages}{2697--2706}.
\newblock
\showISBNx{978-1-4503-1015-4}
\urldef\tempurl%
\url{https://doi.org/10.1145/2207676.2208664}
\showDOI{\tempurl}


\bibitem[Park et~al\mbox{.}(2019)]%
        {park2019slow_algorithm}
\bibfield{author}{\bibinfo{person}{Joon~Sung Park}, \bibinfo{person}{Rick Barber}, \bibinfo{person}{Alex Kirlik}, {and} \bibinfo{person}{Karrie Karahalios}.} \bibinfo{year}{2019}\natexlab{}.
\newblock \showarticletitle{A Slow Algorithm Improves Users' Assessments of the Algorithm's Accuracy}.
\newblock \bibinfo{journal}{\emph{Proc. ACM Hum.-Comput. Interact.}} \bibinfo{volume}{3}, \bibinfo{number}{CSCW}, Article \bibinfo{articleno}{102} (\bibinfo{date}{nov} \bibinfo{year}{2019}), \bibinfo{numpages}{15}~pages.
\newblock
\urldef\tempurl%
\url{https://doi.org/10.1145/3359204}
\showDOI{\tempurl}


\bibitem[Parsons and Haden(2006)]%
        {parsons2006parsons}
\bibfield{author}{\bibinfo{person}{Dale Parsons} {and} \bibinfo{person}{Patricia Haden}.} \bibinfo{year}{2006}\natexlab{}.
\newblock \showarticletitle{Parson's programming puzzles: a fun and effective learning tool for first programming courses}. In \bibinfo{booktitle}{\emph{Proceedings of the 8th Australasian Conference on Computing Education - Volume 52}} (Hobart, Australia) \emph{(\bibinfo{series}{ACE '06})}. \bibinfo{publisher}{Australian Computer Society, Inc.}, \bibinfo{address}{AUS}, \bibinfo{pages}{157–163}.
\newblock
\showISBNx{1920682341}


\bibitem[Peddycord~III et~al\mbox{.}(2014)]%
        {Peddycord2014EDM}
\bibfield{author}{\bibinfo{person}{Barry Peddycord~III}, \bibinfo{person}{Andrew Hicks}, {and} \bibinfo{person}{Tiffany Barnes}.} \bibinfo{year}{2014}\natexlab{}.
\newblock \showarticletitle{Generating hints for programming problems using intermediate output}. In \bibinfo{booktitle}{\emph{Educational Data Mining 2014}}. Citeseer.
\newblock


\bibitem[Pettit et~al\mbox{.}(2017)]%
        {Pettit2017SIGCSE}
\bibfield{author}{\bibinfo{person}{Raymond~S. Pettit}, \bibinfo{person}{John Homer}, {and} \bibinfo{person}{Roger Gee}.} \bibinfo{year}{2017}\natexlab{}.
\newblock \showarticletitle{Do Enhanced Compiler Error Messages Help Students?: Results Inconclusive.}. In \bibinfo{booktitle}{\emph{Proceedings of the 2017 ACM SIGCSE Technical Symposium on Computer Science Education}} (Seattle, Washington, USA) \emph{(\bibinfo{series}{SIGCSE '17})}. \bibinfo{publisher}{ACM}, \bibinfo{address}{New York, NY, USA}, \bibinfo{pages}{465--470}.
\newblock
\showISBNx{978-1-4503-4698-6}
\urldef\tempurl%
\url{https://doi.org/10.1145/3017680.3017768}
\showDOI{\tempurl}


\bibitem[Prather et~al\mbox{.}(2023)]%
        {prather2023robots}
\bibfield{author}{\bibinfo{person}{James Prather}, \bibinfo{person}{Paul Denny}, \bibinfo{person}{Juho Leinonen}, \bibinfo{person}{Brett~A. Becker}, \bibinfo{person}{Ibrahim Albluwi}, \bibinfo{person}{Michelle Craig}, \bibinfo{person}{Hieke Keuning}, \bibinfo{person}{Natalie Kiesler}, \bibinfo{person}{Tobias Kohn}, \bibinfo{person}{Andrew Luxton-Reilly}, \bibinfo{person}{Stephen MacNeil}, \bibinfo{person}{Andrew Petersen}, \bibinfo{person}{Raymond Pettit}, \bibinfo{person}{Brent~N. Reeves}, {and} \bibinfo{person}{Jaromir Savelka}.} \bibinfo{year}{2023}\natexlab{}.
\newblock \showarticletitle{The Robots Are Here: Navigating the Generative AI Revolution in Computing Education}. In \bibinfo{booktitle}{\emph{Proceedings of the 2023 Working Group Reports on Innovation and Technology in Computer Science Education}} (Turku, Finland) \emph{(\bibinfo{series}{ITiCSE-WGR '23})}. \bibinfo{publisher}{Association for Computing Machinery}, \bibinfo{address}{New York, NY, USA}, \bibinfo{pages}{108–159}.
\newblock
\showISBNx{9798400704055}
\urldef\tempurl%
\url{https://doi.org/10.1145/3623762.3633499}
\showDOI{\tempurl}


\bibitem[Prather et~al\mbox{.}(2017)]%
        {Prather2017ICER}
\bibfield{author}{\bibinfo{person}{James Prather}, \bibinfo{person}{Raymond Pettit}, \bibinfo{person}{Kayla~Holcomb McMurry}, \bibinfo{person}{Alani Peters}, \bibinfo{person}{John Homer}, \bibinfo{person}{Nevan Simone}, {and} \bibinfo{person}{Maxine Cohen}.} \bibinfo{year}{2017}\natexlab{}.
\newblock \showarticletitle{On Novices' Interaction with Compiler Error Messages: A Human Factors Approach}. In \bibinfo{booktitle}{\emph{Proceedings of the 2017 ACM Conference on International Computing Education Research}} (Tacoma, Washington, USA) \emph{(\bibinfo{series}{ICER '17})}. \bibinfo{publisher}{ACM}, \bibinfo{address}{New York, NY, USA}, \bibinfo{pages}{74--82}.
\newblock
\showISBNx{978-1-4503-4968-0}
\urldef\tempurl%
\url{https://doi.org/10.1145/3105726.3106169}
\showDOI{\tempurl}


\bibitem[Prather et~al\mbox{.}(2024)]%
        {prather2024benefits_harms_genai}
\bibfield{author}{\bibinfo{person}{James Prather}, \bibinfo{person}{Brent~N Reeves}, \bibinfo{person}{Juho Leinonen}, \bibinfo{person}{Stephen MacNeil}, \bibinfo{person}{Arisoa~S Randrianasolo}, \bibinfo{person}{Brett~A. Becker}, \bibinfo{person}{Bailey Kimmel}, \bibinfo{person}{Jared Wright}, {and} \bibinfo{person}{Ben Briggs}.} \bibinfo{year}{2024}\natexlab{}.
\newblock \showarticletitle{The Widening Gap: The Benefits and Harms of Generative AI for Novice Programmers}. In \bibinfo{booktitle}{\emph{Proceedings of the 2024 ACM Conference on International Computing Education Research - Volume 1}} (Melbourne, VIC, Australia) \emph{(\bibinfo{series}{ICER '24})}. \bibinfo{publisher}{Association for Computing Machinery}, \bibinfo{address}{New York, NY, USA}, \bibinfo{pages}{469–486}.
\newblock
\showISBNx{9798400704758}
\urldef\tempurl%
\url{https://doi.org/10.1145/3632620.3671116}
\showDOI{\tempurl}


\bibitem[Qian and Lehman(2017)]%
        {qian2017intro_prog_difficulties}
\bibfield{author}{\bibinfo{person}{Yizhou Qian} {and} \bibinfo{person}{James Lehman}.} \bibinfo{year}{2017}\natexlab{}.
\newblock \showarticletitle{Students' Misconceptions and Other Difficulties in Introductory Programming: A Literature Review}.
\newblock \bibinfo{journal}{\emph{ACM Trans. Comput. Educ.}} \bibinfo{volume}{18}, \bibinfo{number}{1}, Article \bibinfo{articleno}{1} (\bibinfo{date}{oct} \bibinfo{year}{2017}), \bibinfo{numpages}{24}~pages.
\newblock
\urldef\tempurl%
\url{https://doi.org/10.1145/3077618}
\showDOI{\tempurl}


\bibitem[Ragonis and Ben-Ari(2005)]%
        {ragonis2005long}
\bibfield{author}{\bibinfo{person}{Noa Ragonis} {and} \bibinfo{person}{Mordechai Ben-Ari}.} \bibinfo{year}{2005}\natexlab{}.
\newblock \showarticletitle{A long-term investigation of the comprehension of OOP concepts by novices}.
\newblock \bibinfo{journal}{\emph{Computer Science Education}} \bibinfo{volume}{15}, \bibinfo{number}{3} (\bibinfo{year}{2005}), \bibinfo{pages}{203--221}.
\newblock
\urldef\tempurl%
\url{https://doi.org/10.1080/08993400500224310}
\showDOI{\tempurl}
\showeprint{https://doi.org/10.1080/08993400500224310}


\bibitem[Risko and Gilbert(2016)]%
        {risko2016cognitive}
\bibfield{author}{\bibinfo{person}{Evan~F Risko} {and} \bibinfo{person}{Sam~J Gilbert}.} \bibinfo{year}{2016}\natexlab{}.
\newblock \showarticletitle{Cognitive offloading}.
\newblock \bibinfo{journal}{\emph{Trends in cognitive sciences}} \bibinfo{volume}{20}, \bibinfo{number}{9} (\bibinfo{year}{2016}), \bibinfo{pages}{676--688}.
\newblock


\bibitem[Robinson(2016)]%
        {robinson2016scratch_to_patch}
\bibfield{author}{\bibinfo{person}{William Robinson}.} \bibinfo{year}{2016}\natexlab{}.
\newblock \showarticletitle{From Scratch to Patch: Easing the Blocks-Text Transition}. In \bibinfo{booktitle}{\emph{Proceedings of the 11th Workshop in Primary and Secondary Computing Education}} (M\"{u}nster, Germany) \emph{(\bibinfo{series}{WiPSCE '16})}. \bibinfo{publisher}{Association for Computing Machinery}, \bibinfo{address}{New York, NY, USA}, \bibinfo{pages}{96–99}.
\newblock
\showISBNx{9781450342230}
\urldef\tempurl%
\url{https://doi.org/10.1145/2978249.2978265}
\showDOI{\tempurl}


\bibitem[Sarkar(2024)]%
        {sarkar2024aiprovocateur}
\bibfield{author}{\bibinfo{person}{Advait Sarkar}.} \bibinfo{year}{2024}\natexlab{}.
\newblock \showarticletitle{{AI Should Challenge, Not Obey}}.
\newblock \bibinfo{journal}{\emph{Commun. ACM}} (\bibinfo{date}{Sept.} \bibinfo{year}{2024}), \bibinfo{numpages}{5}~pages.
\newblock
\showISSN{0001-0782}
\urldef\tempurl%
\url{https://doi.org/10.1145/3649404}
\showDOI{\tempurl}
\newblock
\shownote{Online First}.


\bibitem[Savelka et~al\mbox{.}(2023)]%
        {savelka2023gpt4_pass_csed}
\bibfield{author}{\bibinfo{person}{Jaromir Savelka}, \bibinfo{person}{Arav Agarwal}, \bibinfo{person}{Marshall An}, \bibinfo{person}{Chris Bogart}, {and} \bibinfo{person}{Majd Sakr}.} \bibinfo{year}{2023}\natexlab{}.
\newblock \showarticletitle{Thrilled by Your Progress! Large Language Models (GPT-4) No Longer Struggle to Pass Assessments in Higher Education Programming Courses}. In \bibinfo{booktitle}{\emph{Proceedings of the 2023 ACM Conference on International Computing Education Research - Volume 1}} (Chicago, IL, USA) \emph{(\bibinfo{series}{ICER '23})}. \bibinfo{publisher}{Association for Computing Machinery}, \bibinfo{address}{New York, NY, USA}, \bibinfo{pages}{78–92}.
\newblock
\showISBNx{9781450399760}
\urldef\tempurl%
\url{https://doi.org/10.1145/3568813.3600142}
\showDOI{\tempurl}


\bibitem[Scholl and Kiesler(2024)]%
        {scholl2024novice}
\bibfield{author}{\bibinfo{person}{Andreas Scholl} {and} \bibinfo{person}{Natalie Kiesler}.} \bibinfo{year}{2024}\natexlab{}.
\newblock \showarticletitle{How Novice Programmers Use and Experience ChatGPT when Solving Programming Exercises in an Introductory Course}.
\newblock \bibinfo{journal}{\emph{arXiv preprint arXiv:2407.20792}} (\bibinfo{year}{2024}).
\newblock


\bibitem[Scholl et~al\mbox{.}(2024)]%
        {scholl2024analyzing}
\bibfield{author}{\bibinfo{person}{Andreas Scholl}, \bibinfo{person}{Daniel Schiffner}, {and} \bibinfo{person}{Natalie Kiesler}.} \bibinfo{year}{2024}\natexlab{}.
\newblock \showarticletitle{Analyzing Chat Protocols of Novice Programmers Solving Introductory Programming Tasks with ChatGPT}.
\newblock \bibinfo{journal}{\emph{arXiv preprint arXiv:2405.19132}} (\bibinfo{year}{2024}).
\newblock


\bibitem[Skripchuk et~al\mbox{.}(2023)]%
        {skripchuk2023novice_web_help}
\bibfield{author}{\bibinfo{person}{James Skripchuk}, \bibinfo{person}{Neil Bennett}, \bibinfo{person}{Jeffrey Zhang}, \bibinfo{person}{Eric Li}, {and} \bibinfo{person}{Thomas Price}.} \bibinfo{year}{2023}\natexlab{}.
\newblock \showarticletitle{Analysis of Novices' Web-Based Help-Seeking Behavior While Programming}. In \bibinfo{booktitle}{\emph{Proceedings of the 54th ACM Technical Symposium on Computer Science Education V. 1}} (Toronto ON, Canada) \emph{(\bibinfo{series}{SIGCSE 2023})}. \bibinfo{publisher}{Association for Computing Machinery}, \bibinfo{address}{New York, NY, USA}, \bibinfo{pages}{945–951}.
\newblock
\showISBNx{9781450394314}
\urldef\tempurl%
\url{https://doi.org/10.1145/3545945.3569852}
\showDOI{\tempurl}


\bibitem[Sorva et~al\mbox{.}(2013)]%
        {sorva2013program_vis}
\bibfield{author}{\bibinfo{person}{Juha Sorva}, \bibinfo{person}{Ville Karavirta}, {and} \bibinfo{person}{Lauri Malmi}.} \bibinfo{year}{2013}\natexlab{}.
\newblock \showarticletitle{A Review of Generic Program Visualization Systems for Introductory Programming Education}.
\newblock \bibinfo{journal}{\emph{ACM Trans. Comput. Educ.}} \bibinfo{volume}{13}, \bibinfo{number}{4}, Article \bibinfo{articleno}{15} (\bibinfo{date}{nov} \bibinfo{year}{2013}), \bibinfo{numpages}{64}~pages.
\newblock
\urldef\tempurl%
\url{https://doi.org/10.1145/2490822}
\showDOI{\tempurl}


\bibitem[Spohrer and Soloway(1986)]%
        {Spohrer1986CACM}
\bibfield{author}{\bibinfo{person}{James~C. Spohrer} {and} \bibinfo{person}{Elliot Soloway}.} \bibinfo{year}{1986}\natexlab{}.
\newblock \showarticletitle{Novice mistakes: are the folk wisdoms correct?}
\newblock \bibinfo{journal}{\emph{Commun. ACM}} \bibinfo{volume}{29}, \bibinfo{number}{7} (\bibinfo{date}{July} \bibinfo{year}{1986}), \bibinfo{pages}{624–632}.
\newblock
\showISSN{0001-0782}
\urldef\tempurl%
\url{https://doi.org/10.1145/6138.6145}
\showDOI{\tempurl}


\bibitem[Suzuki et~al\mbox{.}(2017)]%
        {Suzuki2017VLHCC}
\bibfield{author}{\bibinfo{person}{R. Suzuki}, \bibinfo{person}{G. Soares}, \bibinfo{person}{A. Head}, \bibinfo{person}{E. Glassman}, \bibinfo{person}{R. Reis}, \bibinfo{person}{M. Mongiovi}, \bibinfo{person}{L. D'Antoni}, {and} \bibinfo{person}{B. Hartmann}.} \bibinfo{year}{2017}\natexlab{}.
\newblock \showarticletitle{TraceDiff: Debugging unexpected code behavior using trace divergences}. In \bibinfo{booktitle}{\emph{2017 IEEE Symposium on Visual Languages and Human-Centric Computing (VL/HCC)}}. \bibinfo{pages}{107--115}.
\newblock
\showISSN{1943-6106}
\urldef\tempurl%
\url{https://doi.org/10.1109/VLHCC.2017.8103457}
\showDOI{\tempurl}


\bibitem[Tamang et~al\mbox{.}(2021)]%
        {tamang2021socratic}
\bibfield{author}{\bibinfo{person}{Lasang~Jimba Tamang}, \bibinfo{person}{Zeyad Alshaikh}, \bibinfo{person}{Nisrine~Ait Khayi}, \bibinfo{person}{Priti Oli}, {and} \bibinfo{person}{Vasile Rus}.} \bibinfo{year}{2021}\natexlab{}.
\newblock \showarticletitle{A Comparative Study of Free Self-Explanations and Socratic Tutoring Explanations for Source Code Comprehension}. In \bibinfo{booktitle}{\emph{Proceedings of the 52nd ACM Technical Symposium on Computer Science Education}} (Virtual Event, USA) \emph{(\bibinfo{series}{SIGCSE '21})}. \bibinfo{publisher}{Association for Computing Machinery}, \bibinfo{address}{New York, NY, USA}, \bibinfo{pages}{219–225}.
\newblock
\showISBNx{9781450380621}
\urldef\tempurl%
\url{https://doi.org/10.1145/3408877.3432423}
\showDOI{\tempurl}


\bibitem[Tankelevitch et~al\mbox{.}(2024)]%
        {tankelevitch2024metacognitive_demands}
\bibfield{author}{\bibinfo{person}{Lev Tankelevitch}, \bibinfo{person}{Viktor Kewenig}, \bibinfo{person}{Auste Simkute}, \bibinfo{person}{Ava~Elizabeth Scott}, \bibinfo{person}{Advait Sarkar}, \bibinfo{person}{Abigail Sellen}, {and} \bibinfo{person}{Sean Rintel}.} \bibinfo{year}{2024}\natexlab{}.
\newblock \showarticletitle{The Metacognitive Demands and Opportunities of Generative AI}. In \bibinfo{booktitle}{\emph{Proceedings of the CHI Conference on Human Factors in Computing Systems}} (Honolulu, HI, USA) \emph{(\bibinfo{series}{CHI '24})}. \bibinfo{publisher}{Association for Computing Machinery}, \bibinfo{address}{New York, NY, USA}, Article \bibinfo{articleno}{680}, \bibinfo{numpages}{24}~pages.
\newblock
\showISBNx{9798400703300}
\urldef\tempurl%
\url{https://doi.org/10.1145/3613904.3642902}
\showDOI{\tempurl}


\bibitem[Tew and Guzdial(2011)]%
        {tew2011fcs1}
\bibfield{author}{\bibinfo{person}{Allison~Elliott Tew} {and} \bibinfo{person}{Mark Guzdial}.} \bibinfo{year}{2011}\natexlab{}.
\newblock \showarticletitle{The FCS1: a language independent assessment of CS1 knowledge}. In \bibinfo{booktitle}{\emph{Proceedings of the 42nd ACM Technical Symposium on Computer Science Education}} (Dallas, TX, USA) \emph{(\bibinfo{series}{SIGCSE '11})}. \bibinfo{publisher}{Association for Computing Machinery}, \bibinfo{address}{New York, NY, USA}, \bibinfo{pages}{111–116}.
\newblock
\showISBNx{9781450305006}
\urldef\tempurl%
\url{https://doi.org/10.1145/1953163.1953200}
\showDOI{\tempurl}


\bibitem[Thompson et~al\mbox{.}(2008)]%
        {thompson2008bloom_cs}
\bibfield{author}{\bibinfo{person}{Errol Thompson}, \bibinfo{person}{Andrew Luxton-Reilly}, \bibinfo{person}{Jacqueline~L. Whalley}, \bibinfo{person}{Minjie Hu}, {and} \bibinfo{person}{Phil Robbins}.} \bibinfo{year}{2008}\natexlab{}.
\newblock \showarticletitle{Bloom's taxonomy for CS assessment}. In \bibinfo{booktitle}{\emph{Proceedings of the Tenth Conference on Australasian Computing Education - Volume 78}} (Wollongong, NSW, Australia) \emph{(\bibinfo{series}{ACE '08})}. \bibinfo{publisher}{Australian Computer Society, Inc.}, \bibinfo{address}{AUS}, \bibinfo{pages}{155–161}.
\newblock
\showISBNx{9781920682590}


\bibitem[Venables et~al\mbox{.}(2009)]%
        {venables2009tracing_writing_followup}
\bibfield{author}{\bibinfo{person}{Anne Venables}, \bibinfo{person}{Grace Tan}, {and} \bibinfo{person}{Raymond Lister}.} \bibinfo{year}{2009}\natexlab{}.
\newblock \showarticletitle{A closer look at tracing, explaining and code writing skills in the novice programmer}. In \bibinfo{booktitle}{\emph{Proceedings of the Fifth International Workshop on Computing Education Research Workshop}} (Berkeley, CA, USA) \emph{(\bibinfo{series}{ICER '09})}. \bibinfo{publisher}{Association for Computing Machinery}, \bibinfo{address}{New York, NY, USA}, \bibinfo{pages}{117–128}.
\newblock
\showISBNx{9781605586151}
\urldef\tempurl%
\url{https://doi.org/10.1145/1584322.1584336}
\showDOI{\tempurl}


\bibitem[Vieira et~al\mbox{.}(2017)]%
        {vieira2017selfexplanation}
\bibfield{author}{\bibinfo{person}{Camilo Vieira}, \bibinfo{person}{Alejandra~J. Magana}, \bibinfo{person}{Michael~L. Falk}, {and} \bibinfo{person}{R.~Edwin Garcia}.} \bibinfo{year}{2017}\natexlab{}.
\newblock \showarticletitle{Writing In-Code Comments to Self-Explain in Computational Science and Engineering Education}.
\newblock \bibinfo{journal}{\emph{ACM Trans. Comput. Educ.}} \bibinfo{volume}{17}, \bibinfo{number}{4}, Article \bibinfo{articleno}{17} (\bibinfo{date}{aug} \bibinfo{year}{2017}), \bibinfo{numpages}{21}~pages.
\newblock
\urldef\tempurl%
\url{https://doi.org/10.1145/3058751}
\showDOI{\tempurl}


\bibitem[Vygotsky(1978)]%
        {vygotsky1978zone}
\bibfield{author}{\bibinfo{person}{Lev~S. Vygotsky}.} \bibinfo{year}{1978}\natexlab{}.
\newblock \bibinfo{booktitle}{\emph{Mind in Society: The Development of Higher Psychological Processes}}.
\newblock \bibinfo{publisher}{Harvard University Press}, \bibinfo{address}{Cambridge, MA}.
\newblock
\showISBNx{9780674576292}


\bibitem[Wang et~al\mbox{.}(2021)]%
        {wang2021learn_barriers_code_example}
\bibfield{author}{\bibinfo{person}{Wengran Wang}, \bibinfo{person}{Archit Kwatra}, \bibinfo{person}{James Skripchuk}, \bibinfo{person}{Neeloy Gomes}, \bibinfo{person}{Alexandra Milliken}, \bibinfo{person}{Chris Martens}, \bibinfo{person}{Tiffany Barnes}, {and} \bibinfo{person}{Thomas Price}.} \bibinfo{year}{2021}\natexlab{}.
\newblock \showarticletitle{Novices' Learning Barriers When Using Code Examples in Open-Ended Programming}. In \bibinfo{booktitle}{\emph{Proceedings of the 26th ACM Conference on Innovation and Technology in Computer Science Education V. 1}} (Virtual Event, Germany) \emph{(\bibinfo{series}{ITiCSE '21})}. \bibinfo{publisher}{Association for Computing Machinery}, \bibinfo{address}{New York, NY, USA}, \bibinfo{pages}{394–400}.
\newblock
\showISBNx{9781450382144}
\urldef\tempurl%
\url{https://doi.org/10.1145/3430665.3456370}
\showDOI{\tempurl}


\bibitem[Watson et~al\mbox{.}(2012)]%
        {Watson2012ICWBL}
\bibfield{author}{\bibinfo{person}{Christopher Watson}, \bibinfo{person}{Frederick~WB Li}, {and} \bibinfo{person}{Jamie~L Godwin}.} \bibinfo{year}{2012}\natexlab{}.
\newblock \showarticletitle{BlueFix: using crowd-sourced feedback to support programming students in error diagnosis and repair}. In \bibinfo{booktitle}{\emph{International Conference on Web-Based Learning}}. Springer, \bibinfo{pages}{228--239}.
\newblock


\bibitem[Wei et~al\mbox{.}(2022)]%
        {wei2022chain_of_thought}
\bibfield{author}{\bibinfo{person}{Jason Wei}, \bibinfo{person}{Xuezhi Wang}, \bibinfo{person}{Dale Schuurmans}, \bibinfo{person}{Maarten Bosma}, \bibinfo{person}{Fei Xia}, \bibinfo{person}{Ed Chi}, \bibinfo{person}{Quoc~V Le}, \bibinfo{person}{Denny Zhou}, {et~al\mbox{.}}} \bibinfo{year}{2022}\natexlab{}.
\newblock \showarticletitle{Chain-of-thought prompting elicits reasoning in large language models}.
\newblock \bibinfo{journal}{\emph{Advances in neural information processing systems}}  \bibinfo{volume}{35} (\bibinfo{year}{2022}), \bibinfo{pages}{24824--24837}.
\newblock


\bibitem[Weinman et~al\mbox{.}(2021)]%
        {weinman2021faded_parsons}
\bibfield{author}{\bibinfo{person}{Nathaniel Weinman}, \bibinfo{person}{Armando Fox}, {and} \bibinfo{person}{Marti~A. Hearst}.} \bibinfo{year}{2021}\natexlab{}.
\newblock \showarticletitle{Improving Instruction of Programming Patterns with Faded Parsons Problems}. In \bibinfo{booktitle}{\emph{Proceedings of the 2021 CHI Conference on Human Factors in Computing Systems}} (Yokohama, Japan) \emph{(\bibinfo{series}{CHI '21})}. \bibinfo{publisher}{Association for Computing Machinery}, \bibinfo{address}{New York, NY, USA}, Article \bibinfo{articleno}{53}, \bibinfo{numpages}{4}~pages.
\newblock
\showISBNx{9781450380966}
\urldef\tempurl%
\url{https://doi.org/10.1145/3411764.3445228}
\showDOI{\tempurl}


\bibitem[White et~al\mbox{.}(2024)]%
        {white2024ai_software_design}
\bibfield{author}{\bibinfo{person}{Jules White}, \bibinfo{person}{Sam Hays}, \bibinfo{person}{Quchen Fu}, \bibinfo{person}{Jesse Spencer-Smith}, {and} \bibinfo{person}{Douglas~C Schmidt}.} \bibinfo{year}{2024}\natexlab{}.
\newblock \bibinfo{booktitle}{\emph{Chatgpt prompt patterns for improving code quality, refactoring, requirements elicitation, and software design}}.
\newblock \bibinfo{publisher}{Springer Nature Switzerland}, \bibinfo{address}{Cham}, \bibinfo{pages}{71--108}.
\newblock
\showISBNx{978-3-031-55642-5}
\urldef\tempurl%
\url{https://doi.org/10.1007/978-3-031-55642-5_4}
\showDOI{\tempurl}


\bibitem[Wing(2006)]%
        {wing2006computational_thinking}
\bibfield{author}{\bibinfo{person}{Jeannette~M. Wing}.} \bibinfo{year}{2006}\natexlab{}.
\newblock \showarticletitle{Computational thinking}.
\newblock \bibinfo{journal}{\emph{Commun. ACM}} \bibinfo{volume}{49}, \bibinfo{number}{3} (\bibinfo{date}{mar} \bibinfo{year}{2006}), \bibinfo{pages}{33–35}.
\newblock
\showISSN{0001-0782}
\urldef\tempurl%
\url{https://doi.org/10.1145/1118178.1118215}
\showDOI{\tempurl}


\bibitem[Xu et~al\mbox{.}(2024)]%
        {xiaotong2024jamplate}
\bibfield{author}{\bibinfo{person}{Xiaotong~(Tone) Xu}, \bibinfo{person}{Jiayu Yin}, \bibinfo{person}{Catherine Gu}, \bibinfo{person}{Jenny Mar}, \bibinfo{person}{Sydney Zhang}, \bibinfo{person}{Jane~L. E}, {and} \bibinfo{person}{Steven~P. Dow}.} \bibinfo{year}{2024}\natexlab{}.
\newblock \showarticletitle{Jamplate: Exploring LLM-Enhanced Templates for Idea Reflection}. In \bibinfo{booktitle}{\emph{Proceedings of the 29th International Conference on Intelligent User Interfaces}} (Greenville, SC, USA) \emph{(\bibinfo{series}{IUI '24})}. \bibinfo{publisher}{Association for Computing Machinery}, \bibinfo{address}{New York, NY, USA}, \bibinfo{pages}{907–921}.
\newblock
\showISBNx{9798400705083}
\urldef\tempurl%
\url{https://doi.org/10.1145/3640543.3645196}
\showDOI{\tempurl}


\bibitem[Yan et~al\mbox{.}(2024)]%
        {litao_chi2024_ivie}
\bibfield{author}{\bibinfo{person}{Litao Yan}, \bibinfo{person}{Alyssa Hwang}, \bibinfo{person}{Zhiyuan Wu}, {and} \bibinfo{person}{Andrew Head}.} \bibinfo{year}{2024}\natexlab{}.
\newblock \showarticletitle{Ivie: Lightweight Anchored Explanations of Just-Generated Code}. In \bibinfo{booktitle}{\emph{Proceedings of the CHI Conference on Human Factors in Computing Systems}} (Honolulu, HI, USA) \emph{(\bibinfo{series}{CHI '24})}. \bibinfo{publisher}{Association for Computing Machinery}, \bibinfo{address}{New York, NY, USA}, Article \bibinfo{articleno}{140}, \bibinfo{numpages}{15}~pages.
\newblock
\showISBNx{9798400703300}
\urldef\tempurl%
\url{https://doi.org/10.1145/3613904.3642239}
\showDOI{\tempurl}


\bibitem[Yilmaz and {Karaoglan Yilmaz}(2023)]%
        {yilmaz2023augmented}
\bibfield{author}{\bibinfo{person}{Ramazan Yilmaz} {and} \bibinfo{person}{Fatma~Gizem {Karaoglan Yilmaz}}.} \bibinfo{year}{2023}\natexlab{}.
\newblock \showarticletitle{Augmented intelligence in programming learning: Examining student views on the use of ChatGPT for programming learning}.
\newblock \bibinfo{journal}{\emph{Computers in Human Behavior: Artificial Humans}} \bibinfo{volume}{1}, \bibinfo{number}{2} (\bibinfo{year}{2023}), \bibinfo{pages}{100005}.
\newblock
\showISSN{2949-8821}
\urldef\tempurl%
\url{https://doi.org/10.1016/j.chbah.2023.100005}
\showDOI{\tempurl}


\bibitem[Zastudil et~al\mbox{.}(2023)]%
        {zastudil2023generative}
\bibfield{author}{\bibinfo{person}{Cynthia Zastudil}, \bibinfo{person}{Magdalena Rogalska}, \bibinfo{person}{Christine Kapp}, \bibinfo{person}{Jennifer Vaughn}, {and} \bibinfo{person}{Stephen MacNeil}.} \bibinfo{year}{2023}\natexlab{}.
\newblock \showarticletitle{Generative AI in Computing Education: Perspectives of Students and Instructors}. In \bibinfo{booktitle}{\emph{2023 IEEE Frontiers in Education Conference (FIE)}}. \bibinfo{pages}{1--9}.
\newblock
\urldef\tempurl%
\url{https://doi.org/10.1109/FIE58773.2023.10343467}
\showDOI{\tempurl}


\bibitem[Zhongxiu~Liu and Barnes(2017)]%
        {liu2017debugging_game}
\bibfield{author}{\bibinfo{person}{Andrew~Hicks Zhongxiu~Liu, Rui~Zhi} {and} \bibinfo{person}{Tiffany Barnes}.} \bibinfo{year}{2017}\natexlab{}.
\newblock \showarticletitle{Understanding problem solving behavior of 6–8 graders in a debugging game}.
\newblock \bibinfo{journal}{\emph{Computer Science Education}} \bibinfo{volume}{27}, \bibinfo{number}{1} (\bibinfo{year}{2017}), \bibinfo{pages}{1--29}.
\newblock
\urldef\tempurl%
\url{https://doi.org/10.1080/08993408.2017.1308651}
\showDOI{\tempurl}


\bibitem[Zingaro et~al\mbox{.}(2018)]%
        {zingaro2018identifying}
\bibfield{author}{\bibinfo{person}{Daniel Zingaro}, \bibinfo{person}{Cynthia Taylor}, \bibinfo{person}{Leo Porter}, \bibinfo{person}{Michael Clancy}, \bibinfo{person}{Cynthia Lee}, \bibinfo{person}{Soohyun Nam~Liao}, {and} \bibinfo{person}{Kevin~C. Webb}.} \bibinfo{year}{2018}\natexlab{}.
\newblock \showarticletitle{Identifying Student Difficulties with Basic Data Structures}. In \bibinfo{booktitle}{\emph{Proceedings of the 2018 ACM Conference on International Computing Education Research}} (Espoo, Finland) \emph{(\bibinfo{series}{ICER '18})}. \bibinfo{publisher}{Association for Computing Machinery}, \bibinfo{address}{New York, NY, USA}, \bibinfo{pages}{169–177}.
\newblock
\showISBNx{9781450356282}
\urldef\tempurl%
\url{https://doi.org/10.1145/3230977.3231005}
\showDOI{\tempurl}


\end{thebibliography}

\end{document}